\documentclass[12pt]{article}
\usepackage{amssymb,amsmath,pstricks,pst-plot, amsfonts}
\usepackage{latexsym}
\usepackage{graphicx}
\usepackage{float}
\usepackage{pstricks}
\usepackage{verbatim}
\usepackage{color}
\usepackage{subfig}
\usepackage{enumerate}
\usepackage{authblk} 
\usepackage{appendix}

\begin{document}
\date{}
\title{Time-delayed feedback control of unstable periodic orbits near a subcritical Hopf bifurcation}
\author[1]{G. Brown}
\author[2]{C.M. Postlethwaite}
\author[1,3]{M. Silber}
\affil[1]{Department of Engineering Sciences and Applied Mathematics, Northwestern University, Evanston, Illinois 60208, USA}
\affil[2]{Department of Mathematics, University of Auckland, Private Bag 92019, Auckland, New Zealand}
\affil[3]{Northwestern Institute on Complex Systems, Northwestern University, Evanston, IL 60208}
\maketitle

\begin{abstract}
We show that Pyragas delayed feedback control can stabilize an unstable periodic orbit (UPO) that arises from a generic subcritical Hopf bifurcation of a stable equilibrium in an $n$-dimensional dynamical system. 
This extends results of Fiedler {\it et al.} [{\it PRL} {\bf 98}, 114101 (2007)], who demonstrated that such feedback control can stabilize the UPO associated with a two-dimensional subcritical Hopf normal form. 
Pyragas feedback requires an appropriate choice of a feedback gain matrix for stabilization, as well as knowledge of the period of the targeted UPO.  We apply feedback in the directions tangent to the two-dimensional center manifold.  We parameterize the feedback gain by a modulus and a phase angle, and give explicit formulae for choosing these two parameters given the period of the UPO in a neighborhood of the bifurcation point. 
We show, first heuristically, and then rigorously by a center manifold reduction for delay differential equations, that the stabilization mechanism involves a highly degenerate Hopf bifurcation problem that is induced by the time-delayed feedback. 
When the feedback gain modulus reaches a threshold for stabilization, {\it both} of the genericity assumptions associated with a two-dimensional Hopf bifurcation are violated: the eigenvalues of the linearized problem do not cross the imaginary axis as the bifurcation parameter is varied, and the real part of the cubic coefficient of the normal form vanishes. Our analysis of this degenerate bifurcation problem reveals two qualitatively distinct cases when unfolded in a two-parameter plane. In each case, Pyragas-type feedback successfully stabilizes the branch of small-amplitude UPOs in a neighborhood of the original bifurcation point, provided that the phase angle satisfies a certain restriction.
\end{abstract}

\section{Introduction}
Chaotic attractors typically possess a dense set of unstable periodic orbits (UPOs).  This form of phase space skeleton was exploited in a control scheme developed by Ott, Grebogi and Yorke~\cite{OGY} in the 1990s. Their approach provided a method for stabilizing targeted UPOs of chaotic systems for application in both numerical simulations and laboratory experiments. This method spawned the development of a number of related and alternative control schemes, with similar goals, which can be utilized in systems for which strictly periodic behavior is attractive~\cite{Handbook}.  One of these schemes, which has been especially well investigated and tested, was proposed by Pyragas~\cite{Pyragas1992}. 

Pyragas control exploits the symmetry of a periodic orbit in a natural way by providing, in its simplest realization, additive feedback in the form 
\begin{equation}
\mathbf{b} = \mathbf{K} \Bigl[\mathbf{x}(t-\tau) - \mathbf{x}(t)\Bigr].
\nonumber 
\end{equation}
Here $\mathbf{x}(t)\in \mathbb{R}^n$ is the state vector of the dynamical system at time $t$, $\tau$ is the period of the targeted UPO, and $\mathbf{K}$ is a constant $n\times n$ feedback gain matrix.  The scheme is manifestly noninvasive, since the feedback vanishes when the system reaches the $\tau$-periodic target state. Setting aside the difficult questions related to basins of attraction, there are then just two key ingredients to the successful implementation of this approach: the period $\tau$ of the targeted UPO is needed, and the feedback gain matrix $\mathbf{K}$ needs to guarantee stabilization.  Only to the extent that an appropriate choice of $\mathbf{K}$ is required does the method rely on detailed knowledge, beyond the period, of the structure of the UPO in phase space.  For a review of the extensive literature on applications of Pyragas feedback, including successful experimental implementations, see~\cite{Pyragas2006}. 

This paper is motivated by the question of  how to choose the feedback gain $\mathbf{K}$ in Pyragas control to ensure that it will be effective. We focus on a simple, generic mechanism for the creation of an unstable periodic orbit: the subcritical Hopf bifurcation of a stable equilibrium.   Other generic mechanisms for creating UPOs in dynamical systems include homoclinic bifurcations, saddle-node (or fold) bifurcations of limit cycles, saddle-node bifurcations of fixed points on an invariant circle, and period-doubling bifurcations~\cite{Strogatz}.  There have been a number of successful demonstrations of Pyragas control of periodic orbits destabilized through a period-doubling bifurcation (see, for instance,~\cite{Pyragasperdoub}).  Postlethwaite~\cite{Claire} has shown that Pyragas-type feedback can stabilize a UPO arising from a subcritical bifurcation from a robust heteroclinic cycle in a three-dimensional system of equivariant ordinary differential equations.  Interestingly, Pyragas feedback works in this case even though the period of the targeted orbit, and hence the time-delay, diverges as the heteroclinic bifurcation point is approached. 
Fiedler~{\it et al.}~\cite{Fiedlerfold} have investigated an example of Pyragas control  that stabilizes a circular limit cycle ({\it i.e.} a ``rotating wave") near a fold bifurcation in a planar system of ordinary differential equations with $O(2)$-symmetry, and successfully applied this to a higher-dimensional model taken from nonlinear optics that possesses a similar rotational symmetry.

The first example that demonstrated the successful stabilization of a UPO arising from a subcritical  Hopf bifurcation was given in~\cite{Fiedler}, and further analyzed in~\cite{Just}, with an experimental implementation described in~\cite{loewenichbennerjust}. In these papers, the authors added Pyragas feedback directly to the Hopf normal form: 
\begin{equation}
{dz(t)\over dt}=(\lambda+i)z(t)+(1+i\gamma)|z(t)|^2z(t)+b_0e^{i\beta}\Bigl[(z(t-\tau)-z(t)\Bigr]. 
\label{eq:normalform}
\end{equation}
Here $\tau\equiv {2\pi\over\ 1 -\gamma\lambda}$ is the period of the UPO, and the complex number $b_0e^{i\beta}$ plays the role of the feedback gain matrix $\mathbf{K}$.  A beauty of this simple example is that it represents  a rare instance in which solutions of a nonlinear delay differential equation can be computed analytically in closed form, and their bifurcations can be studied with comparable finesse. Specifically, 
using methods of bifurcation theory, the authors were able to understand the mechanism for stabilization in this example. For instance, they showed that the feedback control leads to additional delay-induced Hopf bifurcations of the equilibrium $z=0$, and consequently it is possible to change the equilibrium's stability so that the original {\it subcritical} Hopf bifurcation to the UPO turns into a {\it supercritical} bifurcation to a stable periodic orbit. An important contribution of the Fiedler {\it et al.}~\cite{Fiedler} paper was that it also provided a counterexample to a published claim~\cite{Nakajima} that Pyragas control is impossible when the UPO has an odd number of real positive Floquet multipliers greater than one. In the ten years between the published claims of the odd number limitation~\cite{Nakajima} and the first counterexample to it~\cite{Fiedler}, a number of modifications of the Pyragas control scheme were developed. One of these,  based on introducing an additional unstable direction via the controller,  was proposed in order to stabilize UPOs created by a subcritical Hopf bifurcation~\cite{PyragasBenner, PyragasController}, including an example applied to the Lorenz equations~\cite{PyragasLorenz}.
 
Recently an analysis of the Lorenz equations with the standard Pyragas feedback provided a second example of stabilization of a UPO resulting from a subcritical Hopf bifurcation. Specifically, Postlethwaite and Silber~\cite{PS} demonstrated that 
 the stabilization mechanism identified by Fiedler {\it et al.}~\cite{Fiedler} can also apply to UPOs in higher-dimensional systems, provided that the feedback gain matrix is chosen correctly. The strategy they outlined is to add feedback of the type investigated in~\cite{Fiedler,Just} in the directions tangent to the center manifold of the uncontrolled Lorenz system.  Stabilization of the UPOs is then possible over a broad range of control parameter values. 
 
The reduction of higher-dimensional systems to the two-dimensional normal form near a Hopf bifurcation is a standard procedure~\cite{GH}. Likewise, such systems with additive Pyragas feedback, now infinite-dimensional, can also be reduced to the standard two-dimensional normal form in the vicinity of a Hopf bifurcation, where the parameters of the feedback gain matrix modify the coefficients in the normal form. In the Fiedler example~\cite{Fiedler}, the feedback terms are added directly to the Hopf normal form, but a surprising result of~\cite{PS} was that the same sequence of bifurcations identified in the simpler normal form example also appears in this higher-dimensional example.  This result is generalized further in the current paper.
  
A further motivation for the work we present in this paper is to understand the origin of a particular degenerate Hopf bifurcation problem that acts as the organizing center in both the simple normal form example~\eqref{eq:normalform} and the Lorenz example.  Specifically, we generalize the results of~\cite{PS} by studying an $n$-dimensional system of equations containing a subcritical Hopf bifurcation of a stable equilibrium.  As in the Lorenz example, the gain matrix $\mathbf{K}$ for this system is such that the Pyragas feedback only acts in the directions tangent to center manifold of the uncontrolled system near the Hopf bifurcation point, and in this tangent plane the gain collapses to a $2\times 2$ matrix that is proportional to a rotation matrix. Thus, we consider a family of  gain matrices $\mathbf{K}$ that are parameterized  by a magnitude $b_0$ and a phase $\beta$  ({\it cf.}~\eqref{eq:normalform} written in terms of real variables).

The additive Pyragas feedback results in a delay differential equation.  We use methods of bifurcation theory to show that Pyragas control can stabilize the small-amplitude UPO in a neighborhood of its bifurcation provided $b_{0}$ and $\beta$ are chosen appropriately.   Specifically, our analysis applies in a neighborhood of a threshold value for $b_{0}$, which depends on the phase angle $\beta$, which must lie in a particular interval that we determine.  The interval depends only on the cubic coefficient of the Hopf normal form for the uncontrolled problem.  In particular, we find that this interval for $\beta$ always exists provided that the imaginary part of the cubic coefficient of this normal form is nonzero.  The threshold value for $b_0$ is associated with a highly degenerate Hopf bifurcation of the zero solution of the delay differential equation. Specifically, for this gain modulus, the critical eigenvalue of the linearized problem does not cross the imaginary axis as the bifurcation parameter is varied so the ``nonzero-speed" eigenvalue crossing condition for a generic Hopf bifurcation is violated. Moreover,  a center manifold reduction of the delay differential equation to Hopf normal form reveals that the cubic coefficient is purely imaginary for this threshold value of $b_0$, necessitating that one go to higher order than cubic in any analysis of the bifurcating periodic orbits. The analysis of this degenerate bifurcation problem provides the basis for our claims that Pyragas control can stabilize UPOs that are born from a generic subcritical bifurcation of a stable equilibrium in the uncontrolled problem.  It also explains why the same sequence of bifurcations occur in the normal form example~\eqref{eq:normalform} as in the Lorenz example analyzed in~\cite{PS}. 
 
The remainder of this paper is organized as follows.  Section 2 reviews the stabilization mechanisms identified by Fiedler {\em et al.}~\cite{Fiedler} for the Hopf normal form example and formulates our generalized problem.  Section 3 contains our key results.  It determines the restrictions on $\beta$ for effective stabilization.  It also identifies and analyzes the degenerate bifurcation that acts as an organizing center for the control problem.  Section 4 presents a rigorous center manifold reduction for the delay differential equation, with certain details relegated to an Appendix. It thereby substantiates our heuristic arguments made in Section 3. Section 5 summarizes our findings and discusses some open questions and future directions of research.

\section{Problem Formulation}
In this section we review the mechanism of stabilization identified by Fiedler {\it et al.}~\cite{Fiedler} for the Pyragas-controlled Hopf normal form.  We then formulate the generalized problem that will be studied in this paper: an $n$-dimensional system of differential equations containing a generic subcritical Hopf bifurcation of a stable equilibrium, with Pyragas-type delay terms added only in particular directions.  We use a center manifold reduction for the uncontrolled problem to estimate the period of the UPO, which we use as the delay time $\tau$ for the feedback. 

\subsection{Background of Stabilization Mechanism}
\label{sec:formulation}
Fiedler {\it et al.}~\cite{Fiedler} consider equation~\eqref{eq:normalform}, where $z \in \mathbb{C}$ and parameters $\lambda, \gamma \in~\mathbb{R}$.  The feedback gain is a complex number $b_{0}e^{i\beta}$.  For $b_0=0$ and $z = re^{i\theta}$ we have
\begin{align*}
\dot{r} &= \left(\lambda + r^{2}\right)r, \\
\dot{\theta} &= 1 + \gamma r^{2}.
\end{align*}
Bifurcating unstable periodic orbits, or ``Pyragas orbits", with amplitude $r^{2} = -\lambda$, coexist with the stable trivial equilibrium for $\lambda<0$.  The goal is to stabilize this branch of periodic orbits in a neighborhood of $\lambda=0$ by adding the feedback term ($b_{0} \neq 0$).  The Pyragas orbits have minimal period $\tau = 2 \pi/ (1 - \gamma \lambda)$, which is chosen as the delay time in~\eqref{eq:normalform}.

\begin{figure}[t!]
\centering
\subfloat[][] {\includegraphics[width=6cm]{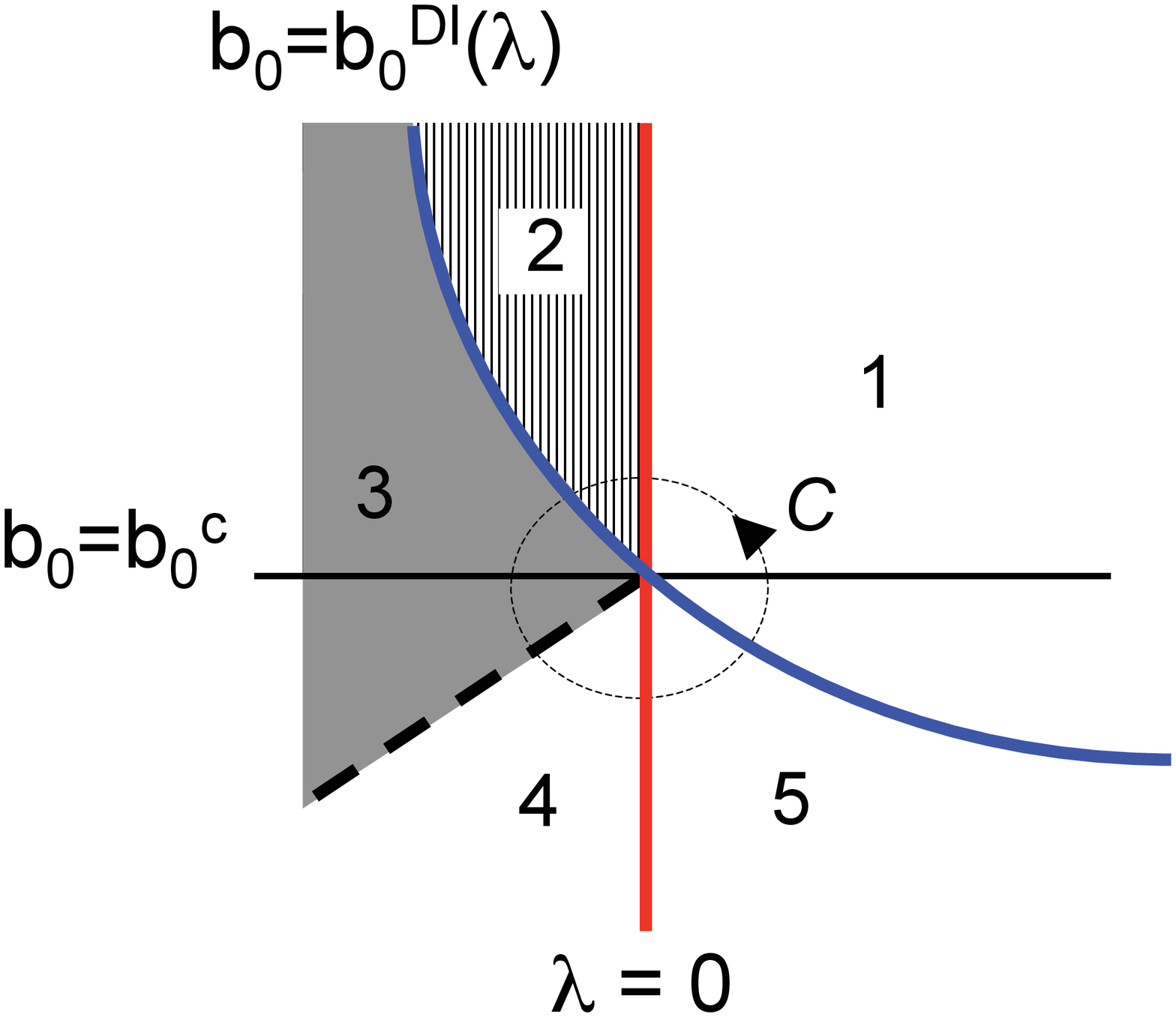}} 
\hspace{0.5cm}
\subfloat[][] {\includegraphics[width=5.5cm]{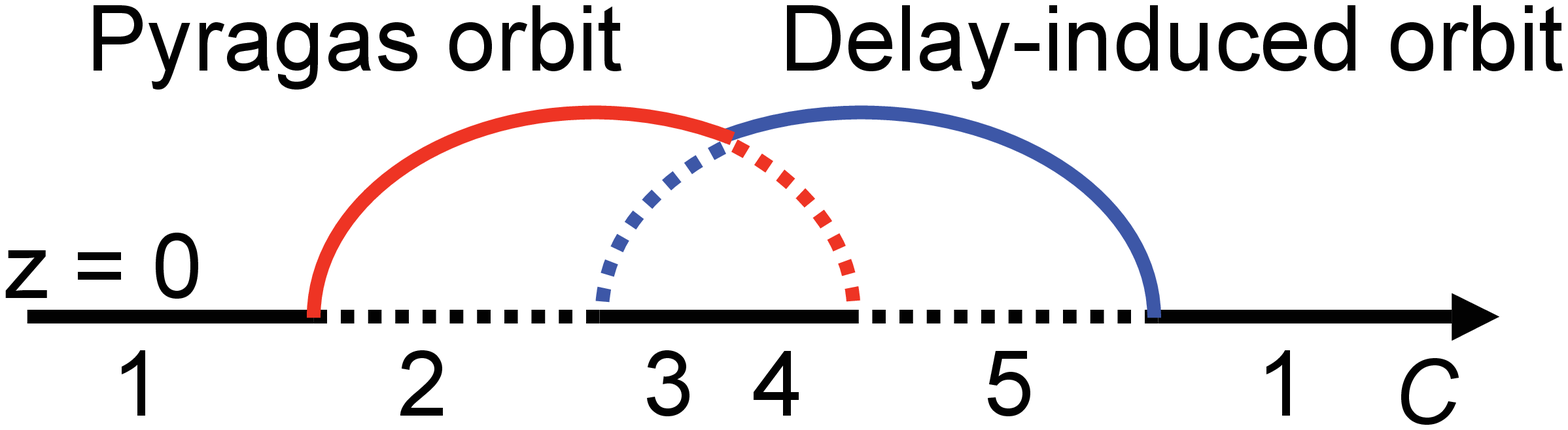}} \\
\caption{(a) Schematic from~\cite{PS} showing Hopf bifurcation curves $\lambda = 0$ (red), $b_{0} = b_{0}^\mathrm{DI}(\lambda)$ (blue), and a transcritical bifurcation curve (dashed line) that divide the $(\lambda,b_{0})$-plane into five regions.  The Pyragas orbit is stable in both the grey shaded and striped regions.  (b)  Schematic showing the bifurcations of the Pyragas and delay-induced periodic orbits as the path $C$ in (a) is traversed counterclockwise about the point $(\lambda, b_{0}) = (0,b_{0}^{c})$. Solid lines indicate stable solutions and dotted lines unstable solutions.}
\label{fig:intro}
\end{figure}

We now summarize the bifurcation structure associated with the $z=0$ solution of~\eqref{eq:normalform} in the $(\lambda, b_{0})$-plane to inform our discussion in subsequent sections. These results, and more details, can be found in~\cite{Fiedler,Just,PS}. Figure~\ref{fig:intro}(a) shows two curves of Hopf bifurcations in the $(\lambda,b_0)$-plane.  One is the original Hopf bifurcation to the Pyragas orbit, which occurs at $\lambda = 0$ for every value of $b_0$. The other Hopf bifurcation is a consequence of the additive delay terms and occurs along the curve $b_{0} = b_{0}^\mathrm{DI}(\lambda)$.  It produces a branch of  {\it delay-induced periodic orbits}, i.e. a periodic orbit that arises due to the addition of delay terms and one for which the feedback does not vanish.  The two bifurcation curves intersect at the point $(\lambda,b_{0})$ = $(0, b_{0}^{c})$.  A curve of transcritical bifurcations also emanates from this point, which acts as the organizing center for the bifurcation structure of the problem.  

Stabilization of the Pyragas orbits involves two bifurcations. Without feedback, the trivial equilbrium is stable for $\lambda<0$ and unstable for $\lambda>0$, and the Hopf bifurcation at $\lambda=0$ is subcritical. However, the delay terms can change the stability of the trivial equilibrium. For $b_0>b_0^c$, we find that the stability of the trivial equilibrium switches to being \emph{unstable} for $\lambda<0$ and \emph{stable} for $\lambda>0$ (in a neighborhood of $\lambda=0$, $b_{0} = b_{0}^{c}$). Since both the location of the Hopf bifurcation at $\lambda=0$, and the location of the Pyragas orbits (in $\lambda<0$) are independent of $b_0$, then the Hopf bifurcation must change criticality from subcritical to supercritical. This in turn means that the Pyragas orbits must now be stable.
The second bifurcation involved in the stabilization mechanism occurs for $\lambda~<~0$. The Pyragas orbit is unstable for small values of $b_{0}$, but as the feedback magnitude is increased, the Pyragas orbit and the delay-induced periodic orbit exchange stability in a transcritical bifurcation.   These mechanisms can both be seen in Fig.~\ref{fig:intro}.

\subsection{Generalized Problem}
The generalized problem is formulated for an $n$-dimensional parameterized system of differential equations 
\begin{equation}
\label{eq:gensys}
\dot{\mathbf{y}} = \mathbf{g}(\mathbf{y},\Lambda), \nonumber
\end{equation}
where $\mathbf{g}: \mathbb{R}^n \times \mathbb{R} \rightarrow \mathbb{R}^{n}$ is $C^{k}$  ($k \geq 5, n \ge 2 $) and $\Lambda$ is a bifurcation parameter.  We assume there is an equilibrium solution branch $y = y_{0}(\Lambda)$ in a neighborhood of $\Lambda = \Lambda_{H}$, which loses stability at $\Lambda = \Lambda_{H}$ as a simple complex conjugate pair of eigenvalues of the linear stability matrix cross the imaginary axis, i.e.\ at a Hopf bifurcation.  We assume the Hopf bifurcation is subcritical, that is, it gives rise to a branch of unstable UPOs which coexist with a stable equilibrium.

Without loss of generality we can introduce shifted variables $\mathbf{x}$ and bifurcation parameter $\lambda$ so that the equilibrium is located at $\mathbf{x = 0}$ for $\Lambda$ in a neighborhood of $\Lambda_{H}$, and $\lambda$ is proportional to $\left(\Lambda - \Lambda_{H}\right)$.  Specifically, we define $\lambda$ so that, in a neighborhood of $\lambda = 0$, $\mathbf{x = 0}$ is stable for $\lambda < 0$ and unstable for $\lambda > 0$.  The bifurcating branch of UPOs exists for $\lambda < 0$.  In terms of the shifted variables, the system is
\begin{equation}
\label{eq:gensysrescaled}
\dot{\mathbf{x}} = \mathbf{f}(\mathbf{x},\lambda), 
\end{equation}
and the Jacobian matrix $D\mathbf{f}(0,\lambda)$ has a pair of complex conjugate eigenvalues $\mu$, $\bar{\mu}$ such that 
\begin{equation}
\mu(\lambda) =  p(\lambda) + i \omega(\lambda), \nonumber
\end{equation}
where
\begin{align}
\label{eq:hopfconditions}
\omega(0) &\equiv \omega_{0} > 0, \nonumber \\
p(0) &=  0,  \nonumber \\
p'(0) &> 0. 
\end{align}
We further assume that the remaining $n-2$ eigenvalues of $D\mathbf{f}(0,0)$ have negative real parts.

\subsection{Delay Time for Feedback}
In order to estimate the period, $T$, of orbits on the branch of UPOs, we perform an (extended) center manifold calculation~\cite{GH} to reduce~\eqref{eq:gensysrescaled} to Hopf normal form 
\begin{equation}
\label{eq:nofbnormform}
\dot{z} = \big( p\left(\lambda \right) + i\omega\left(\lambda\right)\big) z + \big(c\left(\lambda \right) + i d\left(\lambda\right) \big) |z|^2z + \mathcal{O}(|z|^{4}z) \nonumber
\end{equation}
in a neighborhood of $\lambda = 0$. 
In polar coordinates, this is
\begin{align}
\label{eq:nofbpolar}
\dot{r} &= p(\lambda)r + c(\lambda)r^3 + \mathcal{O}(r^5)  \nonumber \\
\dot{\theta} &= \omega(\lambda) + d(\lambda)r^2 + \mathcal{O}(r^4).
\end{align}
We assume that~\eqref{eq:gensysrescaled} depends smoothly on $\lambda$ so that the coefficients can be expanded in Taylor series about $\lambda = 0$:
\begin{align}
\label{eq:taylorseries}
p(\lambda) &= p_{1}\lambda + \mathcal{O}(\lambda^2) \nonumber \\
\omega(\lambda) &= \omega_{0} + \omega_{1}\lambda + \mathcal{O}(\lambda^2)  \nonumber \\
c(\lambda) &= c_{0} + \mathcal{O}(\lambda) \nonumber \\
d(\lambda) &= d_{0} + \mathcal{O}(\lambda).
\end{align}
Then~\eqref{eq:nofbpolar} becomes 
\begin{align}
\label{eq:nofbpolarexpanded}
\dot{r} &= p_{1}\lambda r + c_{0}r^3 + \mathcal{O}(\lambda^{2}r, \lambda r^{3}, r^{5}) \nonumber \\
\dot{\theta} &= \omega_{0} + \omega_{1}\lambda + d_{0}r^{2} + \mathcal{O}(\lambda^{2},\lambda r^{2}, r^{4}).  
\end{align}
Neglecting the higher order terms in~\eqref{eq:nofbpolarexpanded} we find: 
\begin{align}
\label{eq:cubictruncation}
\dot{r} &= p_{1}\lambda r + c_{0}r^3 \nonumber \\
\dot{\theta} &= \omega_{0} + \omega_{1}\lambda + d_{0}r^{2},
\end{align}
and it can be shown \cite{Wiggins} that the dynamics of this truncated normal form are qualitatively unchanged when one considers the influence of the higher order terms provided $p_{1}, c_{0}, \omega_{0} \neq 0$.   

Pyragas orbits exist with amplitude $r^2=-p_{1}\lambda/c_{0}$ for $\lambda < 0$.  (From~\eqref{eq:hopfconditions} we have $p_{1} > 0$, so for a subcritical Hopf bifurcation we must have $c_{0} >$ 0.)  These orbits have period 
\begin{equation}
\label{eq:tau}
T \approx \frac{2\pi}{\omega_{0} + \left(\omega_{1} - p_{1}\gamma \right)\lambda}, 
\end{equation}
where $\gamma = d_{0}/c_{0}$ captures the dependence of the oscillation frequency on the amplitude of oscillations.  Choosing the delay such that $\tau = T$ ensures that the feedback vanishes when the targeted periodic orbit is reached.  Our estimate of $T$ in~\eqref{eq:tau}, which is based on the cubic normal form~\eqref{eq:nofbnormform}, is good through $\mathcal{O}(\lambda)$.   

Next, we add Pyragas-type feedback to~\eqref{eq:gensysrescaled} which gives
\begin{equation}
\label{eq:addfb}
\dot{\mathbf{x}} = \mathbf{f}(\mathbf{x},\lambda) + \mathbf{K} \left(\mathbf{x}(t - \tau) - \mathbf{x}(t) \right), 
\end{equation}
where $\mathbf{K}$ is the constant gain matrix.  As in \cite{PS}, feedback is added only in the directions associated with the linear center eigenspace of~\eqref{eq:gensysrescaled} at the Hopf bifurcation point.  Specifically, after a ($\lambda$-dependent) coordinate transformation and a rescaling of time by the ($\lambda$-dependent) delay $\tau$,~\eqref{eq:addfb} takes the form: 

\begin{align}
\label{eq:origsetup}
\begin{pmatrix}
\dot{x}(t) \\
\dot{y}(t) \\
\dot{\mathbf{w}}(t) \end{pmatrix}   =&  \tau \begin{pmatrix}
\begin{array}{cc}
p(\lambda) & -\omega(\lambda) \\
\omega(\lambda) & p(\lambda)
\end{array}
 & \mathbf{O}_{(2,n-2)} \\ 
\mathbf{O}_{(n-2,2)} & \mathbf{D}(\lambda)
\end{pmatrix}
 \begin{pmatrix} x(t) \\ y(t) \\  \mathbf{w}(t) \\ \end{pmatrix}  + \tau \begin{pmatrix} f_{1}(x(t),y(t),\mathbf{w}(t); \lambda) \\ f_{2}(x(t),y(t),\mathbf{w}(t); \lambda) \\ \mathbf{f}_{d}(x(t),y(t),\mathbf{w}(t); \lambda) \end{pmatrix}  \nonumber \\
 & +  
\tau \begin{pmatrix}
\begin{array}{cc}
b_{0}\cos \beta & -b_{0}\sin \beta \\
b_{0}\sin \beta & b_{0}\cos \beta
\end{array}
 & \mathbf{O}_{(2,n-2)} \\ 
\mathbf{O}_{(n-2,2)} & \mathbf{O}_{(n-2,n-2)}
\end{pmatrix}
 \begin{pmatrix} x(t - 1) - x(t) \\ y(t - 1) - y(t) \\  \mathbf{w}(t - 1) - \mathbf{w}(t) \\ \end{pmatrix}, 
\end{align}
where $\mathbf{w}$ and $\mathbf{f}_{d}$ are $(n-2)$-dimensional column vectors, $\mathbf{O}_{(i,j)}$ is a zero matrix with $i$~rows and $j$ columns, and $\mathbf{D}(\lambda)$ is an $(n-2) \times (n-2)$ matrix in Jordan normal form.  The eigenvalues of $\mathbf{D}(\lambda)$ have negative real part for $\lambda$ sufficiently small.  The feedback depends upon two parameters: an amplitude $b_{0} > 0$ and a phase angle $\beta \in [0,2\pi)$.

\section{Hopf Normal Form for System with Feedback}
\label{sec:delayprob}
We analyze the bifurcation structure of system~\eqref{eq:origsetup} by considering the appropriate two-dimensional Hopf normal form in a neighborhood of $\lambda = 0$.  We show that as $b_{0}$ increases through some critical value $b_{0}^{c}$, the Hopf bifurcation at $\lambda = 0$ changes from subcritical to supercritical, provided the phase angle $\beta$ is chosen appropriately.  Hence there is a range of $b_{0} > b_{0}^{c}$ for which the Pyragas orbit bifurcates stably.  

We show further that the Hopf bifurcation at the point $(\lambda,b_{0}) = (0,b_{0}^{c})$ is degenerate for two reasons: (a) the nonzero eigenvalue crossing condition of the Hopf bifurcation theorem is violated, and (b) the cubic coefficient of the normal form is purely imaginary.  First we perform a linear stability analysis to show that the nonzero crossing condition is violated, which leads to an explicit formula for $b_{0}^{c}$.  Another linear consideration, specifically the requirement that the Pyragas branch bifurcates from a stable equilibrium,  determines restrictions on the parameters $\gamma$ and $\beta$.  In particular, we must assume that $\gamma \neq 0$, and we require that $\beta$ lie within a specified range.   

We use results from the linear analysis, together with information on the Pyragas branch, to argue that the real part of the cubic coefficient of the Hopf normal form also vanishes at $\lambda = 0$, $b_{0} = b_{0}^{c}$.  This result is later substantiated in Section~\ref{sec:cmreduction} (with details in Appendix~B) by a center manifold reduction of the delay differential equation~\eqref{eq:origsetup}.  The degeneracy at cubic order necessitates that quintic terms in the Hopf normal form be retained.  The bifurcation analysis for this problem is performed at the end of this section.  

\subsection{Degeneracy Associated with the Linear Normal Form Coefficient} 
\label{sec:lin}
We first perform a linear stability analysis to determine the feedback magnitude at which the eigenvalue crossing condition is violated. Since the feedback terms in~\eqref{eq:origsetup} only act in the center directions, and all the other coordinates are linearly decaying, we focus our linear stability analysis on the $\dot{x}$ and $\dot{y}$ equations.  In terms of the complex variable $z = x + iy$ we obtain the linear delay differential equation:
\begin{equation}
\dot{z}(t) = \tau(\lambda) \left( p(\lambda) + i\omega(\lambda) \right) z(t) + \tau(\lambda) b_{0}e^{i\beta}\left(z(t-1) - z(t)\right), \nonumber
\end{equation}
with $\beta \in [0, 2\pi)$ and $b_{0} > 0$.  Solutions take the form $z = e^{\eta t}$, where $\eta$ satisfies the characteristic equation $\chi(\eta) = 0$, with
\begin{equation}
\label{eq:chareqn}
\chi(\eta) \equiv \tau(\lambda) \left( p(\lambda) + i\omega(\lambda) \right) + \tau(\lambda) b_{0}e^{i\beta}\left(e^{-\eta} - 1\right) - \eta.
\end{equation}

At $\lambda = 0$, this equation becomes
\begin{equation}
\label{eq:chareqn0}
2 \pi i + \frac{2 \pi}{\omega_{0}} b_{0}e^{i\beta}\left(e^{-\eta} - 1 \right) - \eta = 0, \nonumber
\end{equation}
since $\tau(0) = 2\pi / \omega_{0}$, $p(0) = 0$, and $\omega(0) \equiv \omega_{0}$.  This has a solution $\eta= \eta_{0} = 2\pi i$, independent of $b_{0}$, which is as expected since the original Hopf bifurcation is not affected by the feedback.  

We next evaluate the eigenvalue crossing condition, that is, we consider $\eta$ to be a function of $\lambda$ with $\eta(0) = 2\pi i$, and compute $Re[\eta'(0)]$.  Specifically, we expand $\eta(\lambda)$ about $\lambda = 0$ as
\begin{equation}
\label{eq:etaexp}
\eta(\lambda)=2\pi i+\eta_1\lambda+\mathcal{O}(\lambda^2),
\end{equation}
so $\eta'(0)=\eta_1$.
The Taylor series for $p(\lambda)$ and $\omega(\lambda)$ were given in~\eqref{eq:taylorseries}.  The delay $\tau(\lambda)$,  given by~\eqref{eq:tau}, can be expanded to obtain
\begin{equation}
\label{eq:tauexpansion}
\tau(\lambda) = \tau_{0} + \tau_{1} \lambda + \mathcal{O}(\lambda^{2}), \nonumber
\end{equation}
where 
\begin{align}
\tau_{0} =& \frac{2\pi}{\omega_{0}} > 0 \nonumber \\
\tau_{1} =& \frac{-2\pi\left(\omega_{1} - p_{1}\gamma\right)}{\omega_{0}^{2}}. \nonumber
\end{align}
Equating terms at $\mathcal{O}(\lambda)$ gives
 \begin{equation}
 \label{eq:eta1}
 \eta_1 = \frac{2\pi p_{1}\left(1 + i\gamma\right)}{\omega_{0} + 2 \pi b_{0}e^{i\beta}}.  \nonumber
 \end{equation}
We find that the crossing condition is violated, that is, $Re\left[\eta'(0) \right] = 0$  when
\begin{equation}
\label{eq:b0crit}
b_{0} = b_{0}^{c} \equiv \frac{-\omega_{0}}{2\pi \left(\cos \beta + \gamma \sin \beta \right)}, \hspace{1cm} \beta \neq 0,\pi. 
\end{equation} 
In order to have a positive (finite) value for $b_{0}^{c}$, the phase angle $\beta$ must satisfy the restriction
\begin{equation}
\label{eq:betarestriction}
\cos \beta + \gamma \sin \beta < 0, \hspace{1cm} \beta \neq 0,\pi. 
\end{equation}

Moreover, as discussed in~\cite{Fiedler}, the feedback introduces additional delay-induced instabilities of the $z = 0$ solution of~\eqref{eq:origsetup}.  In order to ensure that the Pyragas branch can bifurcate from a stable solution, we need to ensure that $b_{0}$ is below a $\beta$-dependent cut-off where these additional instabilities set in.  These considerations will determine a more stringent inequality for $\beta$, which can be met provided that $\gamma \neq 0$, where $\gamma \equiv d(0)/c(0)$ in \eqref{eq:nofbnormform}.

We know that as $b_{0} \rightarrow 0$ (i.e. the feedback vanishes), solutions of the characteristic equation~\eqref{eq:chareqn} at $\lambda = 0$ include $\eta_{0} = 2 \pi i$, and a countable set that have real parts tending to $-\infty$~\cite{Elsgolts}.  We determine the value of $b_{0}$ (with $\lambda = 0$) for the onset of the first delay-induced bifurcation of $z = 0$ by seeking solutions $\eta = i \nu$ of~\eqref{eq:chareqn} for $\lambda = 0$.  Specifically, $(\nu, b_{0})$ satisfy 
 \begin{align*}
  \cos(\beta - \nu) &= \cos\beta\\
  b_{0} \left( \sin(\beta - \nu) - \sin\beta \right) &= (\nu - 2\pi)\frac{\omega_{0}}{2\pi}. 
 \end{align*}
 These equations have one solution at $\nu=2\pi$ for all values of $b_0$ (which corresponds to the original Hopf bifurcation), and a sequence $(\nu,b_{0}) = (\nu_{j},b_{0}^{j})$, indexed by $j \in \mathbb{Z}$, where
 \[
 \nu_{j} =2\beta+2\pi j,\quad \hspace{0.5cm} \quad b_{0}^{j} = -\frac{\omega_{0}}{2\pi}\left( \frac{\beta + (j-1)\pi}{\sin\beta}  \right).
 \]
 
 For the Hopf bifurcation with $\nu=2\pi$ to be from a stable equilibrium when $b_0=b_0^c$, we need to ensure that, for each $j$, either $b_0^j>b_0^c$ or $b_0^j<0$. Since $\beta \in (0, 2\pi)$ with $\beta \neq \pi$, this condition is satisfied if
  \begin{equation}
  0 < b_{0}^{c} < b_{0}^{0}, \nonumber
  \end{equation}
  where 
  \begin{equation}
  \label{eq:b0critrepeat}
  b_{0}^{c} = -\frac{\omega_{0}}{2\pi}\left( \frac{1}{\cos\beta + \gamma \sin\beta} \right) 
  \end{equation}
  \begin{equation}
  \label{eq:b0j0}
  b_{0}^{0} = -\frac{\omega_{0}}{2\pi}\left(	\frac{\beta - \pi}{\sin\beta} \right). 
  \end{equation}
Hence we require that $\beta \in (0, 2\pi)$ satisfies
 \begin{equation}
 \label{eq:restriction}
 \cos\beta + \gamma \sin\beta  < \frac{\sin\beta}{\beta - \pi}, \hspace{1cm} \beta \neq \pi.
 \end{equation}  
Note that $\frac{\sin(\beta)}{(\beta - \pi)} < 0$ for $\beta \in (0,2\pi)$, so~\eqref{eq:restriction} automatically ensures that~\eqref{eq:betarestriction} is satisfied. 

{\bf Claim:} If $\gamma \neq 0$, then there exists a value of $\beta$ such that~\eqref{eq:restriction} is satisfied.  
{\bf Proof:} We rewrite~\eqref{eq:restriction} as
 \newcommand{\tb}{\tilde{\beta}}
 \[ -\cos(\beta-\pi)-\gamma\sin(\beta-\pi)<\frac{-\sin(\beta-\pi)}{\beta-\pi}
  \]
  So, equivalently, we must show that there exists some $\tb\in(-\pi,\pi)\setminus{0}$ such that
\begin{equation}
\label{eq:firstcond} 
\cos\tb+\gamma\sin\tb >\frac{\sin\tb}{\tb}, 
\end{equation}
which we can rewrite as 
\begin{equation}
\label{eq:lastcond}
\sqrt{1+\gamma^2}\cos(\tb-\tilde{\beta}_{\mathrm{min}}) >\frac{\sin\tb}{\tb}, 
\end{equation}
where $\tan\tilde{\beta}_{\mathrm{min}}=\gamma$.  (Note, for $\tilde{\beta} = \tilde{\beta}_{\mathrm{min}}$, one obtains the smallest possible value of $b_{0}^{c}$ given by~\eqref{eq:b0crit}).  
Now, observe that $(\sin\tb)/\tb<1$ for  $\tb\in(-\pi,\pi)\setminus{0}$. Since $\sqrt{1+\gamma^2}> 1$, then it is clear that there is an open interval  of $\tb$ for which~\eqref{eq:lastcond}  is satisfied. This interval will include the point where $\cos(\tb-\tilde{\beta}_{\mathrm{min}})=1$, that is, where $\tb=\tilde{\beta}_{\mathrm{min}}=\tan^{-1}(\gamma)$.  $\square$

Note that when $\gamma=0$, condition~\eqref{eq:firstcond} becomes $\cos \tb >\sin \tb /\tb$, or equivalently,
\begin{align*}
\tan\tb&<\tb, \quad \mathrm{for} \quad \tb\in(-\pi,-\pi/2)\cup(0,\pi/2), \\
\tan\tb&>\tb, \quad \mathrm{for} \quad \tb\in(-\pi/2,0)\cup(\pi/2,\pi), 
\end{align*}
which is clearly never satisfied.  If $\tb = \pm \pi/2$, then $(\sin\tb)/\tb = 2/\pi$ and~\eqref{eq:firstcond} is still never satisfied.  Since condition~\eqref{eq:firstcond} does not hold for any $\tb \in (-\pi,\pi)\setminus{0}$ when $\gamma = 0$, the mechanism of stabilization which we discuss in this paper is not valid for $\gamma = 0$.  

Henceforth we assume that $\gamma \neq 0$ and that $\beta$ is chosen to satisfy~\eqref{eq:restriction}.  We focus on a neighborhood of $\lambda = 0, b_{0} = b_{0}^{c}$, expanding the eigenvalue $\eta$ in~\eqref{eq:etaexp} in a two-variable Taylor series in order to zoom in on the behavior near the point $(\lambda, b_{0}) = (0, b_{0}^{c})$: 
\begin{equation}
\label{eq:eta}
\eta(\lambda, \delta) = 2\pi i + \eta_{2}\delta + \eta_{3}\lambda^2 + \eta_{4}\lambda \delta + \eta_{5}\delta^2 + O(\lambda^3, \lambda^{2}\delta, \lambda \delta^{2}, \delta^{3})
\end{equation}
where 
\begin{equation}
\delta \equiv b_{0} - b_{0}^{c}.  \nonumber
\end{equation}
Note that there is no linear term in $\lambda$, since we already know that at $\delta=0$, $\eta'(\lambda)=0$. Also, when $\lambda = 0$,  $\eta = 2\pi i$ is a solution to $\chi(\eta) = 0$ for all~$\delta$, which immediately implies that $\eta_{2} = \eta_{5} = 0$.  We assume, generically, that $Re(\eta_{3}) \neq 0$.  Note that to determine $\eta_{3}$ we would need, among other quantities, the delay time $\tau$ through order $\lambda^{2}$, which cannot be computed using the cubic truncation of the Hopf normal form~\eqref{eq:cubictruncation}.  
 
\begin{figure}[t!]
  \centering
  \subfloat[][$\sigma_{3} < 0$] {\label{fig:sigma3neg} \includegraphics[height=4.5cm]{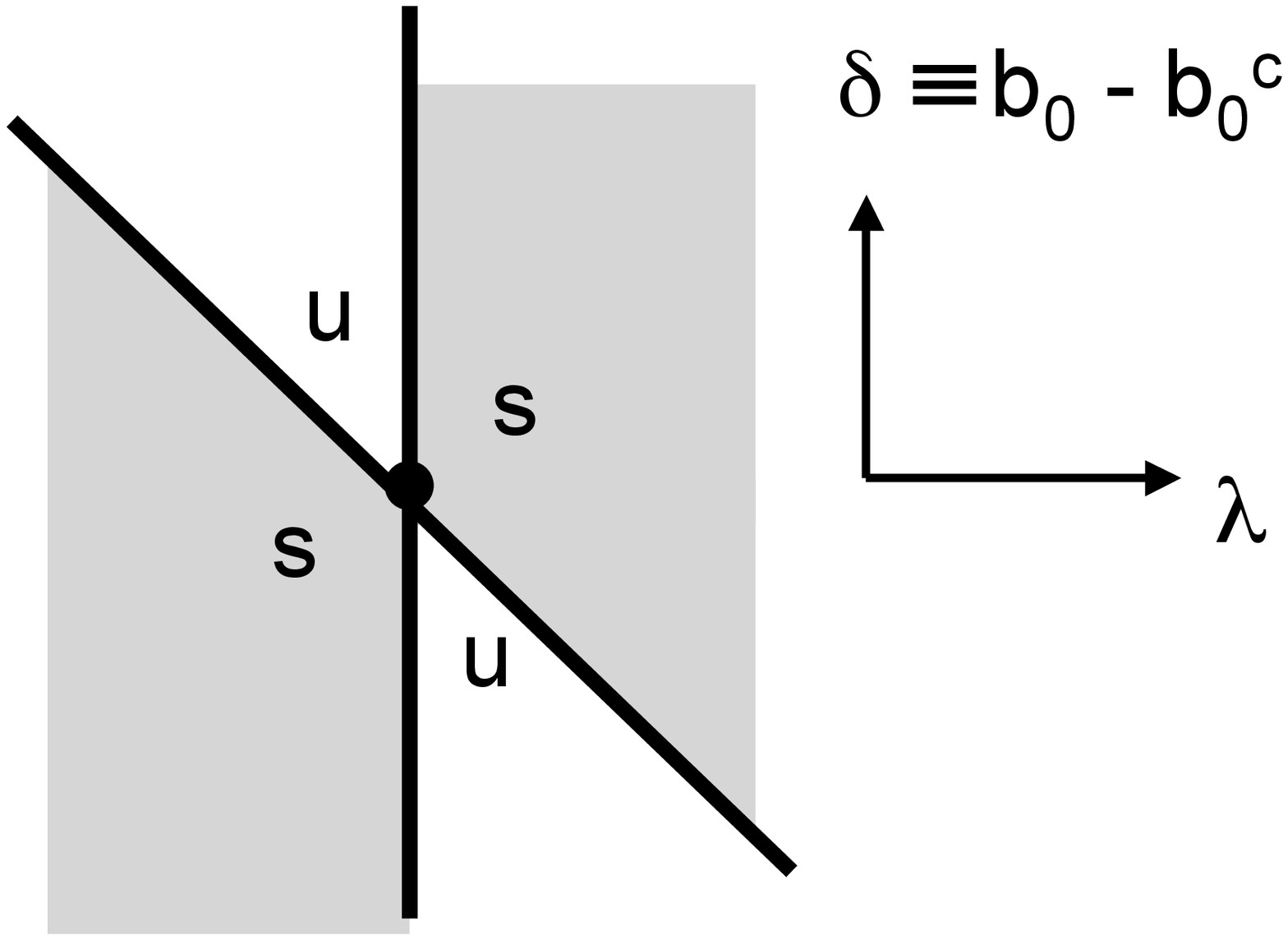}} 
\subfloat[][$\sigma_{3} > 0$] {\label{fig:sigma3pos} \includegraphics[height=4.5cm]{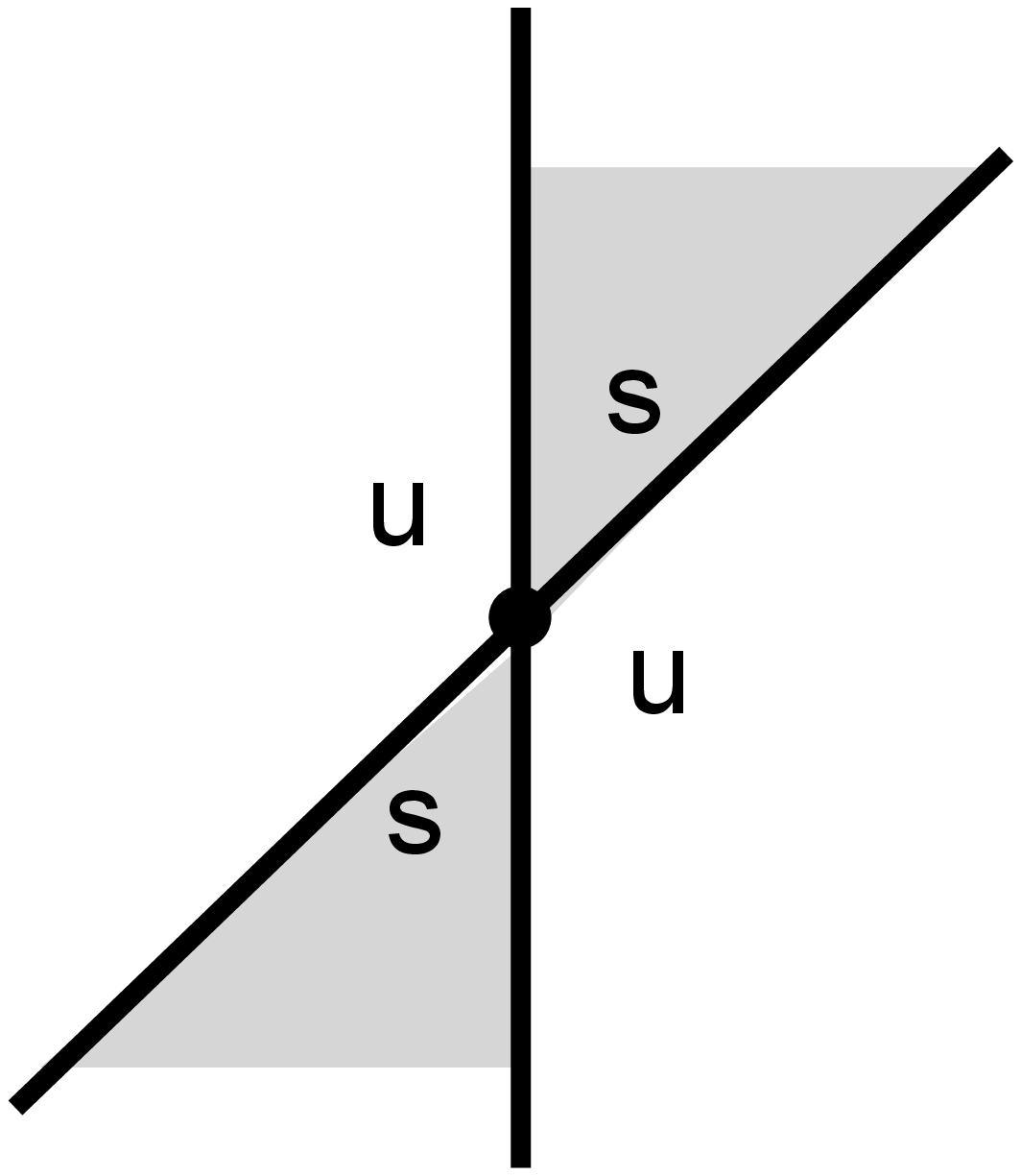}}         
 \caption{Schematic depiction of the stability of the zero equilibrium in the $(\lambda,\delta)$-plane valid in a neighborhood of $\lambda = \delta = 0$ from~\eqref{eq:sigma} when (a) $\sigma_{3} < 0$ and when (b) $\sigma_{3} > 0$.  Stable (unstable) regions are shaded (unshaded) and marked with s (u).}
  \label{fig:lin_stab}
\end{figure}

Substituting~\eqref{eq:eta} into~\eqref{eq:chareqn} and equating terms at $\mathcal{O}(\lambda \delta)$ yields 
\begin{equation}
\label{eq:eta4}
Re(\eta_{4}) = \frac{4 p_{1}\pi^{2}}{\omega_{0}^{2}}\frac{\left(\cos \beta + \gamma \sin \beta \right)^{3}}{(\gamma^2 + 1)\sin^{2}\beta} < 0, 
\end{equation}
where the inequality follows from~\eqref{eq:hopfconditions} and~\eqref{eq:betarestriction}. 
Finally, defining $\sigma(\lambda,\delta) \equiv Re(\eta)$, we have
\begin{equation}
\label{eq:sigma}
\sigma(\lambda,\delta) = \sigma_{3}\lambda^2 + \sigma_{4}\lambda\delta + \mathcal{O}(\lambda^{3},\lambda^{2}\delta, \lambda \delta^{2}) 
\end{equation}
where $\sigma_{4}\equiv Re(\eta_4) < 0$ by~\eqref{eq:eta4}, and $\sigma_3=Re(\eta_3)\neq 0$ (generically).  From this we are able to deduce the arrangement of the regions of stability of the zero equilibrium around the point $(\lambda, \delta) = (0,0)$. The arrangement depends on the sign of $\sigma_{3}$, and the two cases are shown in Fig.~\ref{fig:lin_stab}.

\subsection{Degeneracy Associated with the Cubic Normal Form Coefficient}
The linear stability analysis reveals that the nonzero eigenvalue crossing condition is violated at $b_{0} = b_{0}^{c}$.  We now argue that the cubic coefficient of the normal form equation is purely imaginary at $\delta \equiv b_{0} - b_{0}^{c} = 0$, which is the second degeneracy of the Hopf bifurcation at $(\lambda,\delta) = (0,0)$.  This follows directly, as we now show, from the linear calculations in Section~\ref{sec:lin} and the fact that the existence of the Pyragas orbit is unaffected by the addition of the feedback terms.

According to bifurcation theory for dynamical systems, including delay equations,~\eqref{eq:origsetup} restricted to its center manifold in a neighborhood of the Hopf bifurcation point at $\lambda = 0$ can be converted into normal form via a series of nonlinear, near-identity coordinate transformations  \cite{GH, Hassard, FariaMagalhaes, Qesmi}.  This normal form is
\begin{equation}
\label{eq:cmnormalformcomplex}
\dot{z} = \eta z + k |z|^{2}z + q |z|^{4}z + \mathcal{O}(|z|^{6}z).
\end{equation}
where $z, \eta, k, q \in \mathbb{C}$.  Here we have retained the quintic terms in anticipation of the result that $Re(k) = 0$ at $\lambda = \delta = 0$.  We will assume that $Re(q) \neq 0$, and analyze the quintic truncation of~\eqref{eq:cmnormalformcomplex}.  We demonstrate for a specific numerical example in Appendix~A that the coefficient of the fifth-order term does not vanish, which we expect to be true generically.  Rewriting~\eqref{eq:cmnormalformcomplex} in polar coordinates, truncating the terms above fifth order, and considering only the real part of the equation, we have
\begin{equation}
\label{eq:cmnormalform}
\dot{r} = r \left( \sigma  + k_{R}r^2 + q_{R}r^4 \right),
\end{equation}
where $\sigma = \sigma(\lambda,\delta)$ is the expansion given by~\eqref{eq:sigma} and $k_{R} = k_{R}(\lambda,\delta)$, $q_{R} = q_{R}(\lambda,\delta)$ are the Taylor series expansions for the real parts of the coefficients of the cubic and quintic terms, respectively. 

The zeros of~\eqref{eq:cmnormalform} correspond to the limit cycle solutions on the center manifold of the original problem~\eqref{eq:origsetup} in a neighborhood of $\lambda = 0$.  (Although the periodic orbits are not circular in the original coordinates on the center manifold, the normal form transformation is the coordinate transformation that makes them circular.)  The Pyragas orbit is, by construction, unaffected by the control and must exist as a zero of~\eqref{eq:cmnormalform} for $\lambda < 0$.  From~\eqref{eq:nofbpolar}, we have that the Pyragas orbit satisfies $r^2=-p(\lambda)/c(\lambda)$, for which the first order approximation is $r^2 = -p_{1}\lambda/c_{0}$.  We then factor~\eqref{eq:cmnormalform} to obtain
\begin{equation}
\label{eq:factorednormform}
\dot{r} = r\left( p + cr^{2} \right) \left( a + br^{2} \right),
\end{equation}
where $\sigma = pa, k_{R} = ca + pb$, and $q_{R} = cb$.  Thus
\begin{equation}
\label{eq:kreal}
k_{R} = \frac{c\sigma}{p} + \frac{qp}{c}, \nonumber
\end{equation}
where $\sigma$ is known from the linear problem, and $p,c$ are known from the original uncontrolled Hopf bifurcation and are independent of $\delta$.  

To find a Taylor expansion of  $k_R$ in terms of  $\lambda$ and $\delta$, we substitute the Taylor expansions of $p,c$ (given by~\eqref{eq:taylorseries}) and the Taylor series of $\sigma$ (given by~\eqref{eq:sigma}) to obtain 
\begin{equation}
k_{R}(\lambda, \delta) = \frac{c_{0}\sigma_{3}\lambda}{p_{1}} + \frac{q_{R0}p_{1}\lambda}{c_{0}} + \frac{c_{0}\sigma_{4} \delta}{p_{1}} + O(\lambda^{2}, \lambda \delta, \delta^{2}). \nonumber
\end{equation}
It follows immediately that $k_{R}(0,0) = 0$.  We have thus shown that the real part of the cubic coefficient of the normal form vanishes at $(\lambda, \delta) = (0,0)$.

\subsection{Bifurcation Analysis}
We now analyze~\eqref{eq:taylorseries} in a neighborhood of $(\lambda, \delta) = (0,0)$ with a goal of determining all qualitatively distinct bifurcation diagrams associated with the distinguished bifurcation parameter $\lambda$, and showing that there is a region of parameter space in which the Pyragas orbit is stable. 

Expanding all coefficients in~\eqref{eq:factorednormform} in Taylor series in $\lambda$ and $\delta \equiv b_{0} - b_{0}^{c}$, we have, at leading order, 
\begin{equation}
\label{eq:factored}
\dot{r} = r \left( p_{1}\lambda + c_{0}r^{2} \right)\left( \alpha_{1}\lambda + \alpha_{2}\delta + q_{0}r^{2} \right),
\end{equation}
where 
\begin{align*}
\alpha_{1} &\equiv \frac{\sigma_{3}}{p_{1}}, \nonumber \\
\alpha_{2} &\equiv \frac{\sigma_{4}}{p_{1}} < 0, \nonumber \\
q_{0} &\equiv \frac{q_{R0}}{c_{0}}.
\end{align*}
The inequality for $\alpha_{2}$ follows from the fact that $p_{1} > 0$ and $\sigma_{4} < 0$.  

The Pyragas orbits always exist for $\lambda < 0$.  On the other hand, the delay terms in~\eqref{eq:origsetup} create additional Hopf bifurcations that give rise to delay-induced periodic orbits.  From~\eqref{eq:factored}, we see that the delay-induced periodic orbits exist provided 
\begin{equation}
r^{2} = \frac{-\alpha_{1} \lambda - \alpha_{2} \delta}{q_{0}} > 0, \nonumber
\end{equation} 
from which we find the delay-induced Hopf bifurcation line
\begin{equation}
\label{eq:dinhopfline}
\delta = \delta_{DI} = \frac{-\alpha_{1}\lambda}{\alpha_{2}}
\end{equation}
in the $(\lambda, \delta)$-plane.  If $q_{0} > 0$, the delay-induced periodic orbits exist in the region of the $(\lambda, \delta)$-plane above this line, while if $q_{0} < 0$ they exist in the region below this line.  

A transcritical bifurcation of periodic orbits occurs when the Pyragas and delay-induced periodic orbits have the same amplitude, so that the right-hand side of~\eqref{eq:factored} is a perfect square.  The equation for the line of transcritical bifurcations is
\begin{equation}
\label{eq:tchopfline}
\delta = \delta_{TC} = \left( \frac{p_{1}q_{0} - \alpha_{1}c_{0}}{\alpha_{2}c_{0}} \right) \lambda, 
\end{equation}
with $p_{1}, c_{0} > 0$. Note that~\eqref{eq:dinhopfline} and \eqref{eq:tchopfline} are only correct through~$\mathcal{O}(\lambda)$ and that this bifurcation curve must terminate at $\lambda = 0$ since the Pyragas orbits only exist for $\lambda < 0$.  

The arrangement of the curves~\eqref{eq:dinhopfline} and \eqref{eq:tchopfline} in the $(\lambda, \delta)$-plane gives two qualitatively different cases for the bifurcation structure of~\eqref{eq:factored}, each of which may be divided into three subcases when the distinguished bifurcation parameter $\lambda$ is varied with fixed $\delta$.  Recall that $c_0,p_1>0$ and $\alpha_2<0$; we assume the generic situation where $\alpha_1, q_{0} \neq 0$. If $q_0\alpha_1<0$, then the sign of $p_{1}q_{0} - \alpha_{1}c_{0}$ is determined, but otherwise, it can be of either sign.  (If $p_{1}q_{0} - \alpha_{1}c_{0} = 0$, then the line of transcritical bifurcations occurs at $\delta = \delta_{TC} = 0$.)  The possible cases are thus:
\begin{equation}
\label{tab:cases}
\begin{tabular}{|c|c|c|}
\hline
Case (1): $q_{0} > 0$ &  &  \\ \hline
(a)  & $\alpha_{1} > 0$ & $p_{1}q_{0} - \alpha_{1}c_{0} \geq 0$ \\ \hline
(b) & $\alpha_{1} > 0$ & $p_{1}q_{0} - \alpha_{1}c_{0} \leq 0$ \\ \hline
(c) & $\alpha_{1} < 0$ & $p_{1}q_{0} - \alpha_{1}c_{0} \geq 0$ \\ \hline
Case (2): $q_{0} < 0$ & & \\ \hline
(a)	& $\alpha_{1} > 0$ & $p_{1}q_{0} - \alpha_{1}c_{0} \leq 0$ \\ \hline
(b)	& $\alpha_{1} < 0$ & $p_{1}q_{0} - \alpha_{1}c_{0} \leq 0$ \\ \hline
(c)	& $\alpha_{1} < 0$ & $p_{1}q_{0} - \alpha_{1}c_{0} \geq 0$ \\ \hline    
\end{tabular}
\end{equation}

\begin{figure}[t!]
\centering
\hspace{-1.5cm}
\subfloat[][Case (1a)] {\includegraphics[width=4.5cm]{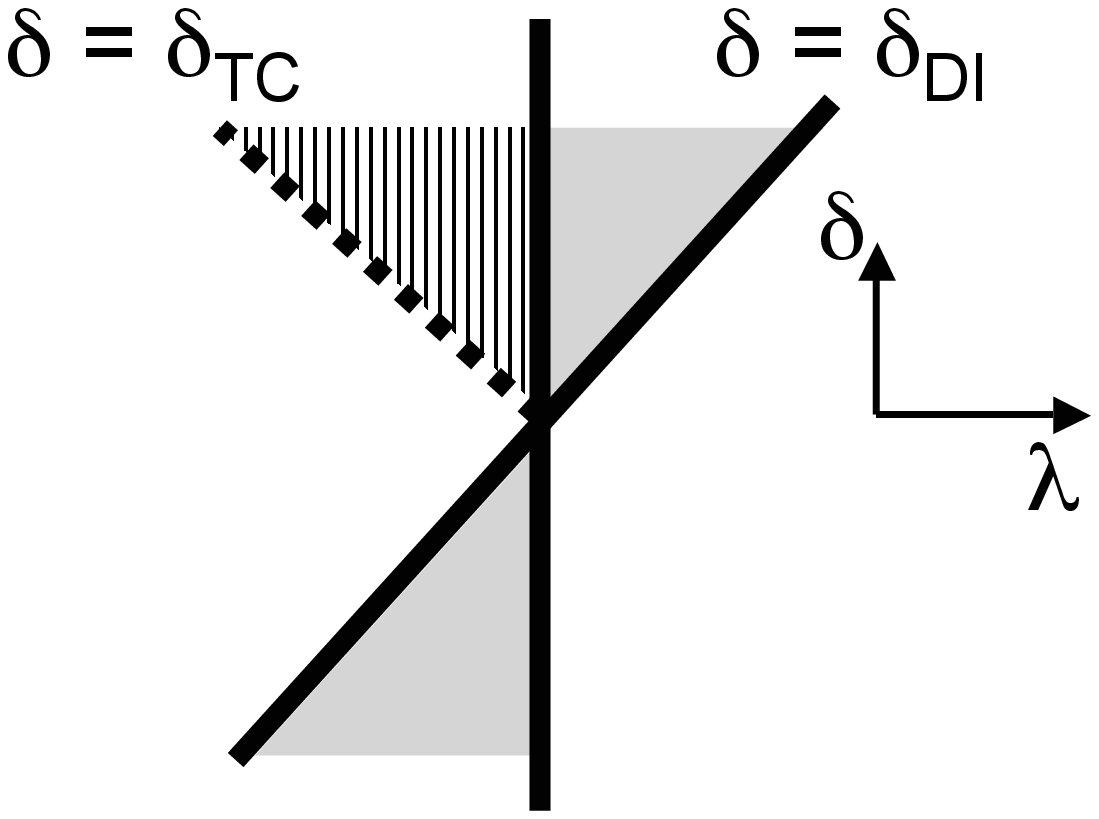}} 
\subfloat[][Case (1b)] {\includegraphics[width=4.5cm]{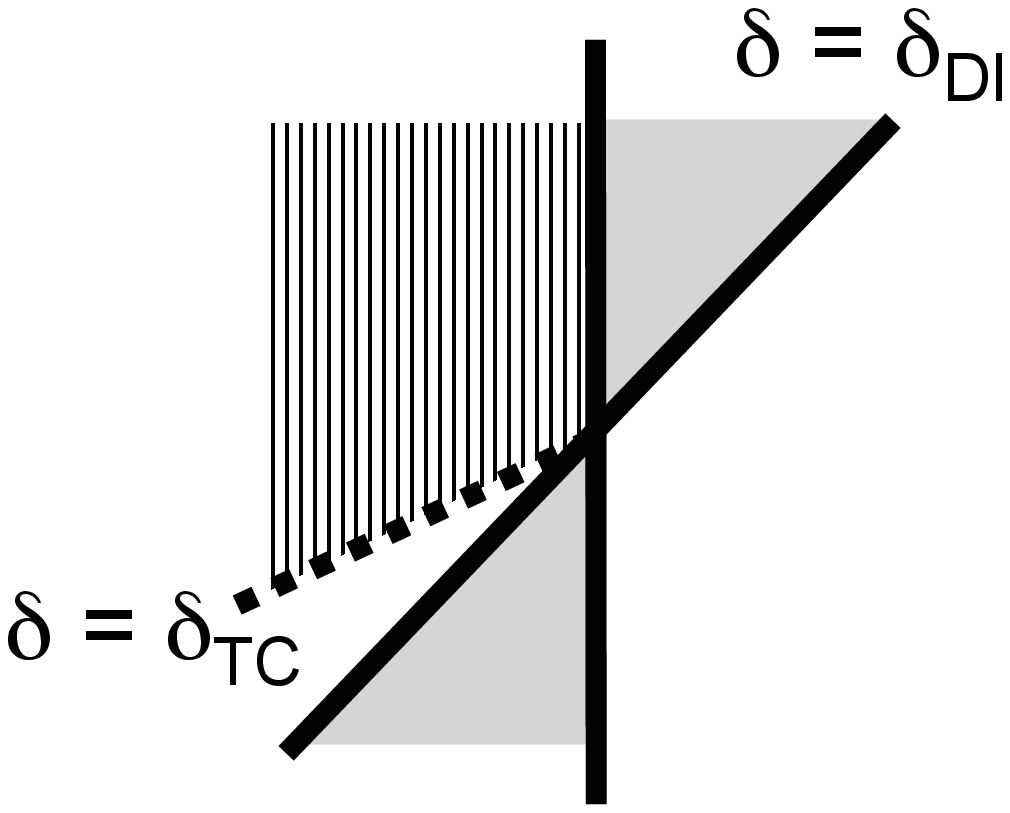}}
\subfloat[][Case (1c)] {\includegraphics[width=4.5cm]{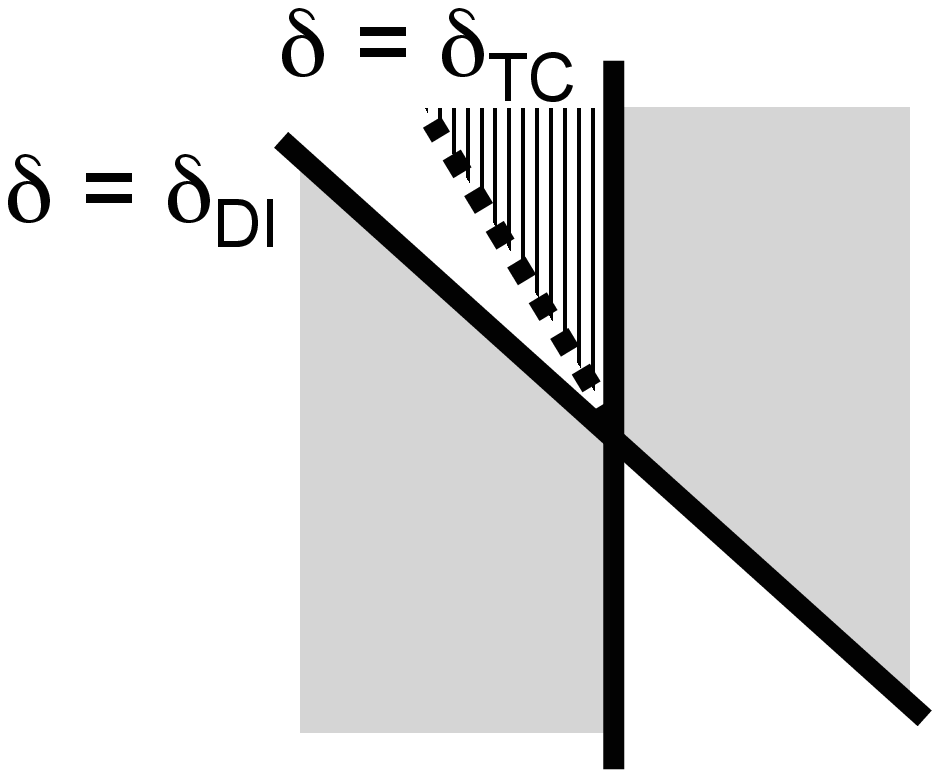}} \\
\hspace{-1.5cm}
\subfloat[][Case (2a)] {\includegraphics[width=4.5cm]{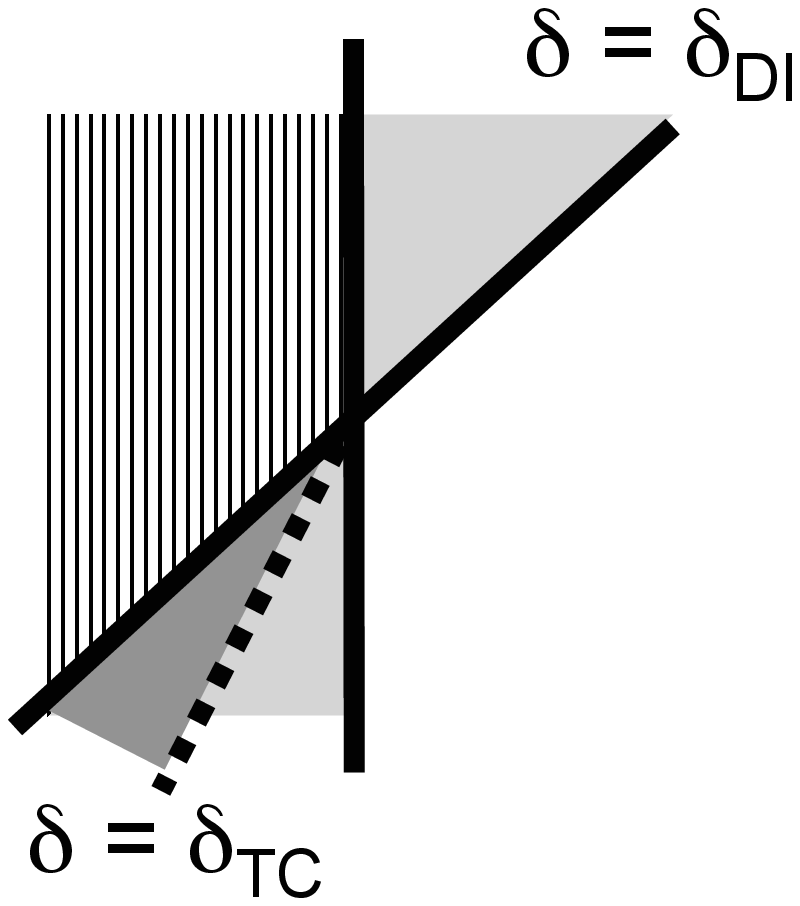}} 
\subfloat[][Case (2b)] {\includegraphics[width=4.5cm]{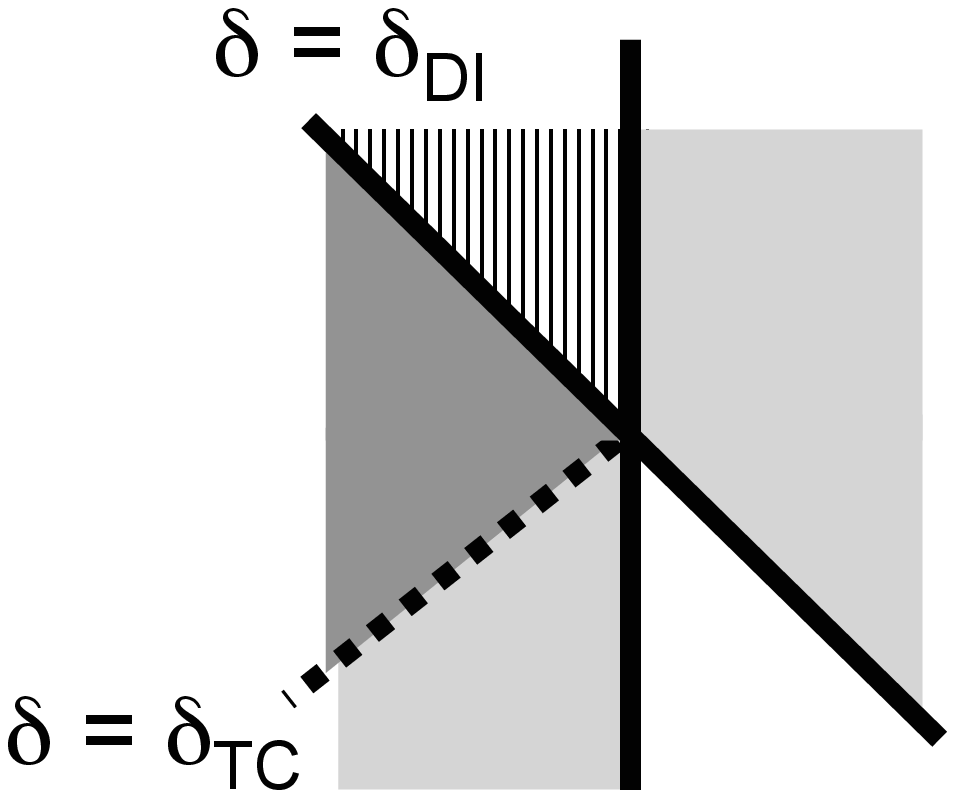}} 
\subfloat[][Case (2c)] {\includegraphics[width=4.5cm]{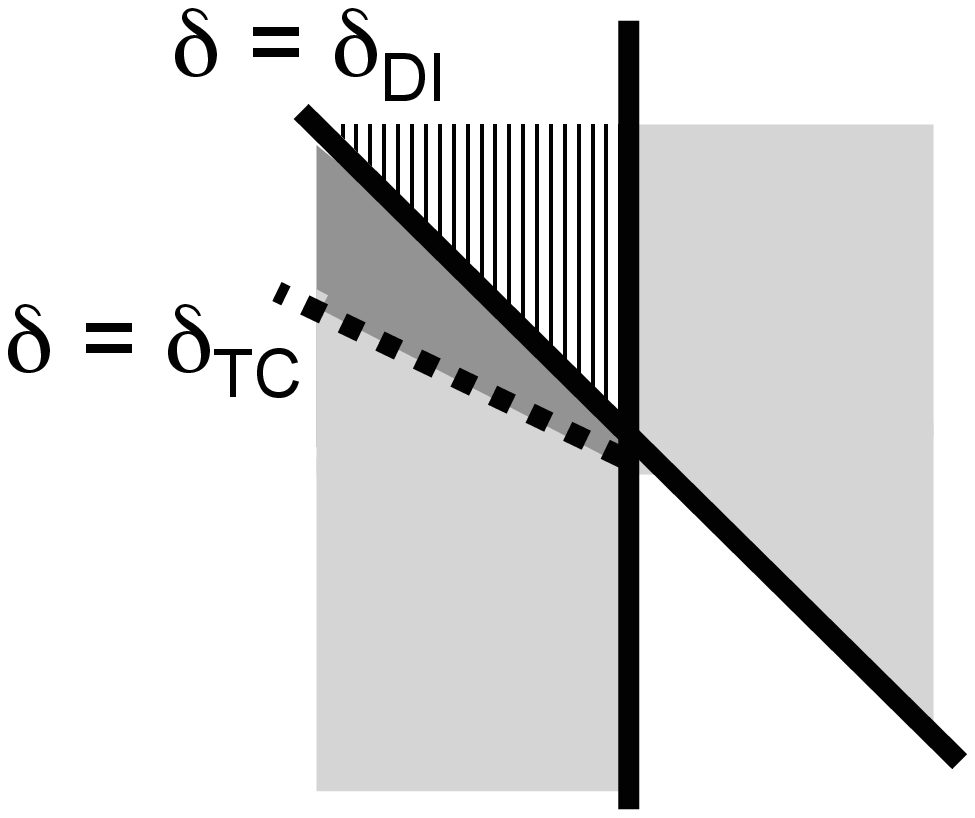}} 
\caption{Bifurcation sets for~\eqref{eq:factored} in the $(\lambda,\delta)$-plane for the qualitatively distinct cases from~\eqref{tab:cases}.  Curves of Hopf  (transcritical) bifurcations are indicated by a solid (dotted) black line; $\delta = \delta_{DI}$ are the delay-induced bifurcation curves and $\delta = \delta_{TC}$ are the transcritical bifurcation curves.  The trivial equilibrium is stable in the light and dark grey shaded regions.  The Pyragas orbit is stable in the dark grey and striped regions.  The delay-induced orbit is stable for $\delta_{DI} < \delta < \delta_{TC}$ in Case 1, and for $\delta < \delta_{TC}$ in Case 2.}
\label{fig:fourcases}
\end{figure}

\begin{figure}[t!]
\centering
\subfloat[][Case (1)] {\label{fig:smooth} \includegraphics[width=6cm]{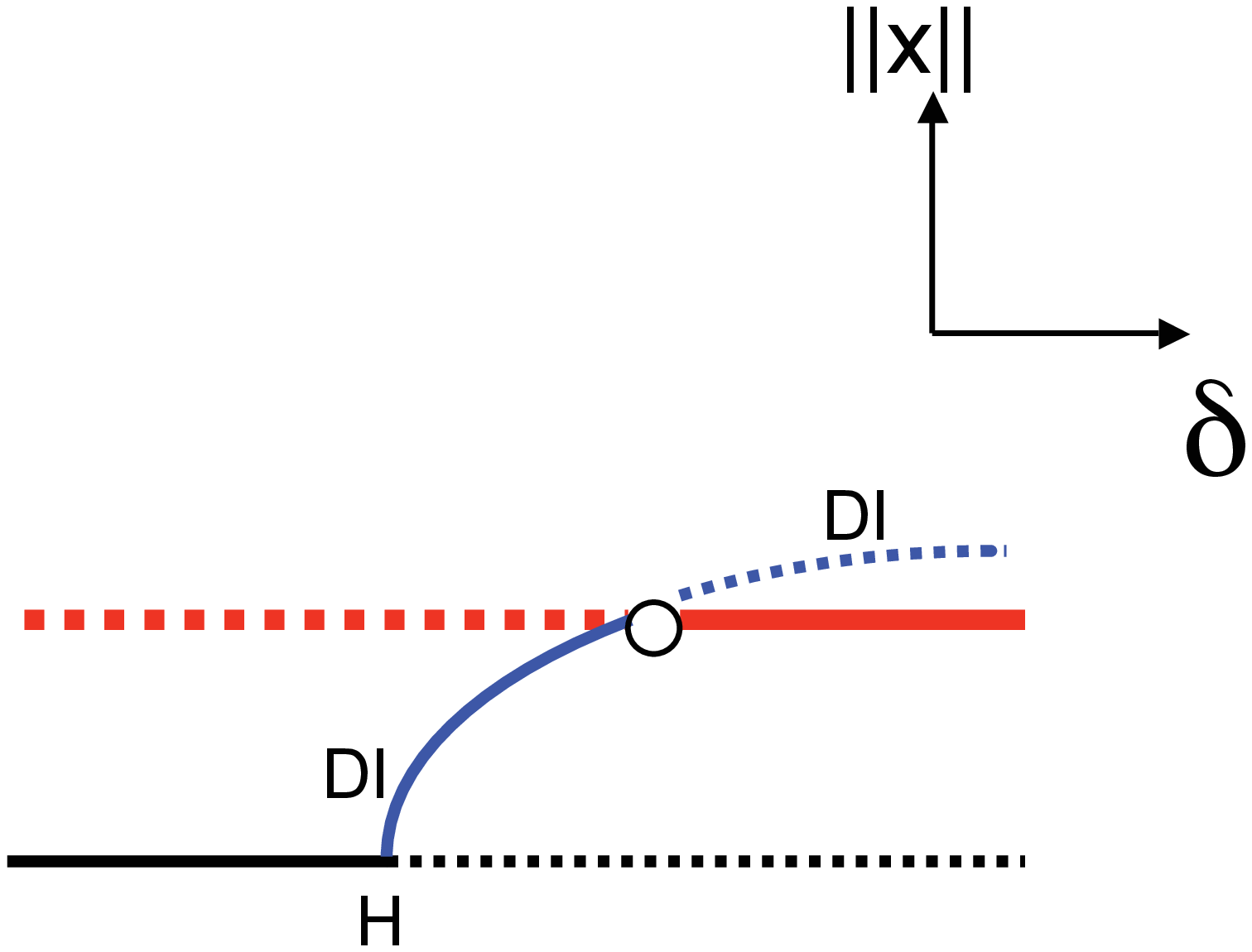}} 
\hspace{-0.5cm}
\subfloat[][Case (2)] {\label{fig:jump} \includegraphics[width=6cm]{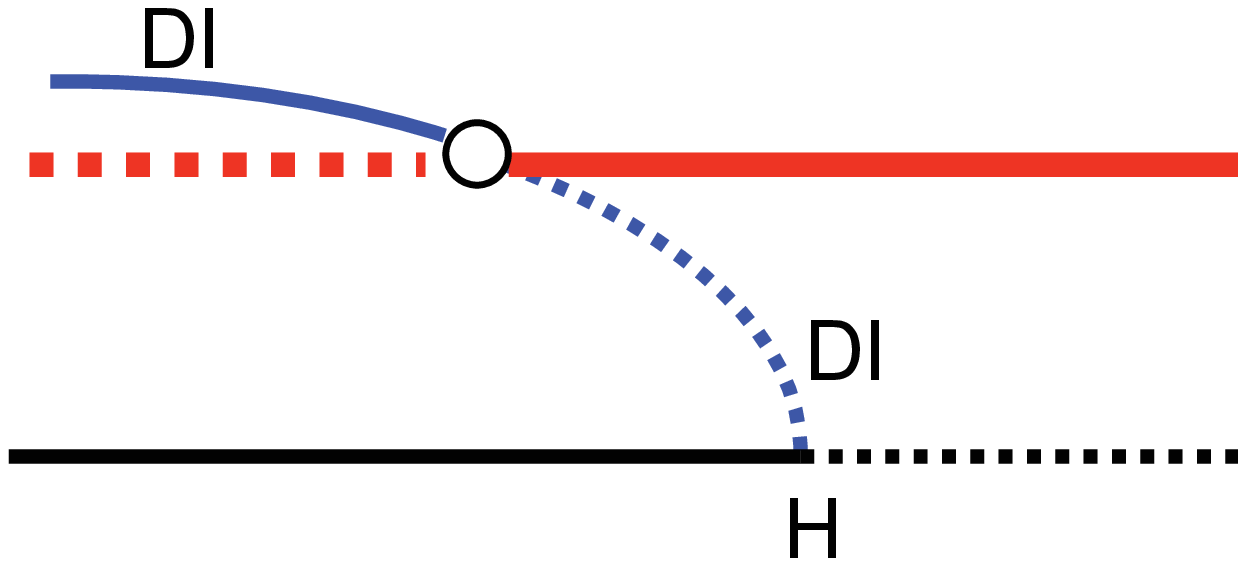}} 
\caption{Bifurcation diagrams indicating the amplitude of the periodic orbits vs. the parameter $\delta \equiv b_{0} - b_{0}^{c}$ for fixed $\lambda < 0$ in~\eqref{eq:factored}.  Solid (dotted) lines indicate stable (unstable) solutions.  The Hopf bifurcation to the delay-induced orbit is indicated by an H, and the transcritical bifurcation is indicated by an open dot.  The delay-induced periodic orbits are labeled DI.  The Pyragas orbits are unlabeled red lines, which are horizontal because their amplitude is independent of $\delta$.}
\label{fig:deltavaries}
\end{figure}

\begin{figure}[h!]
\centering
\hspace{-1.5cm}
\subfloat[][Case (1a)] {\includegraphics[width=5cm]{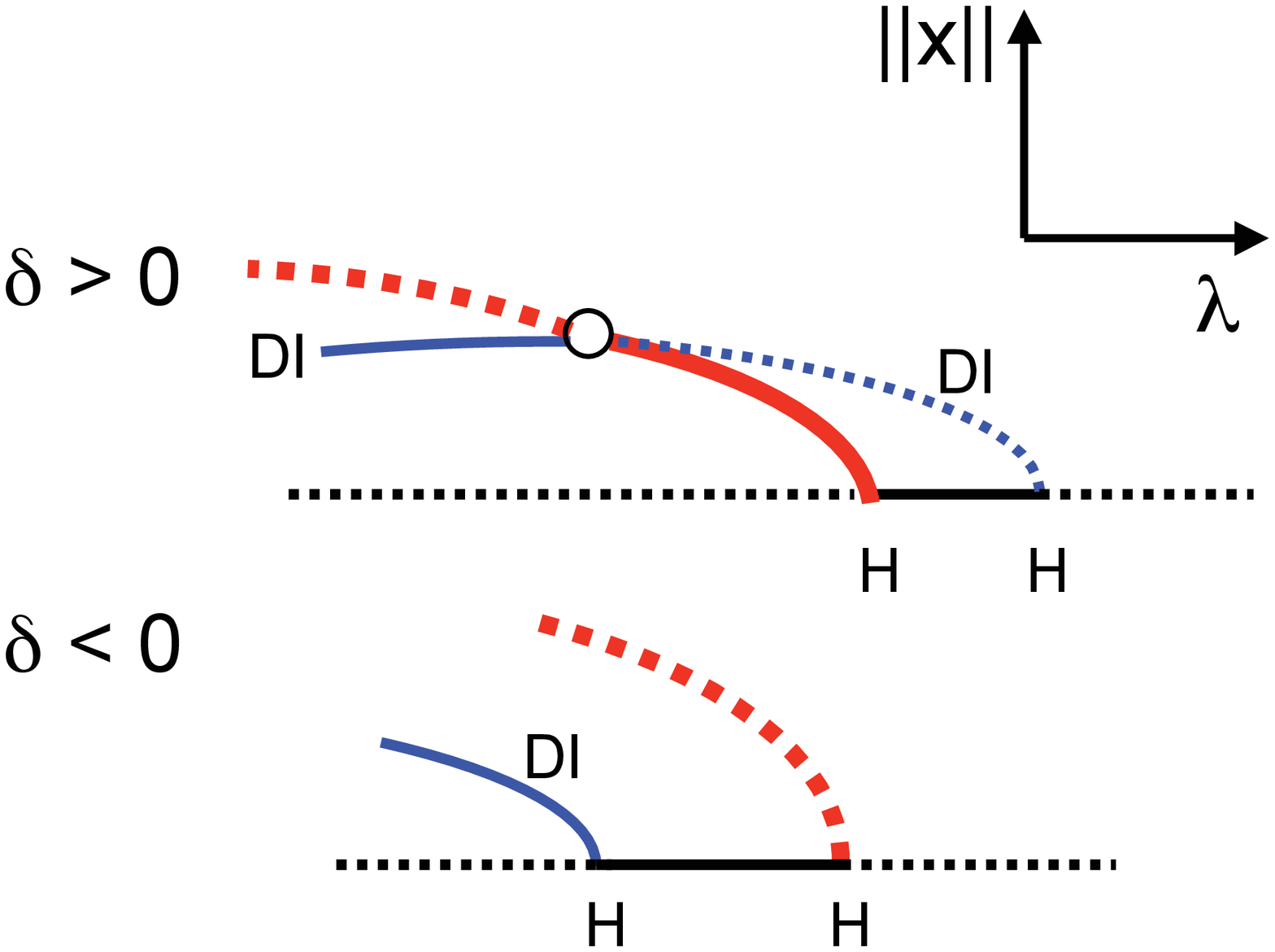}} 
\hspace{-1cm}
\subfloat[][Case (1b)] {\includegraphics[width=5cm]{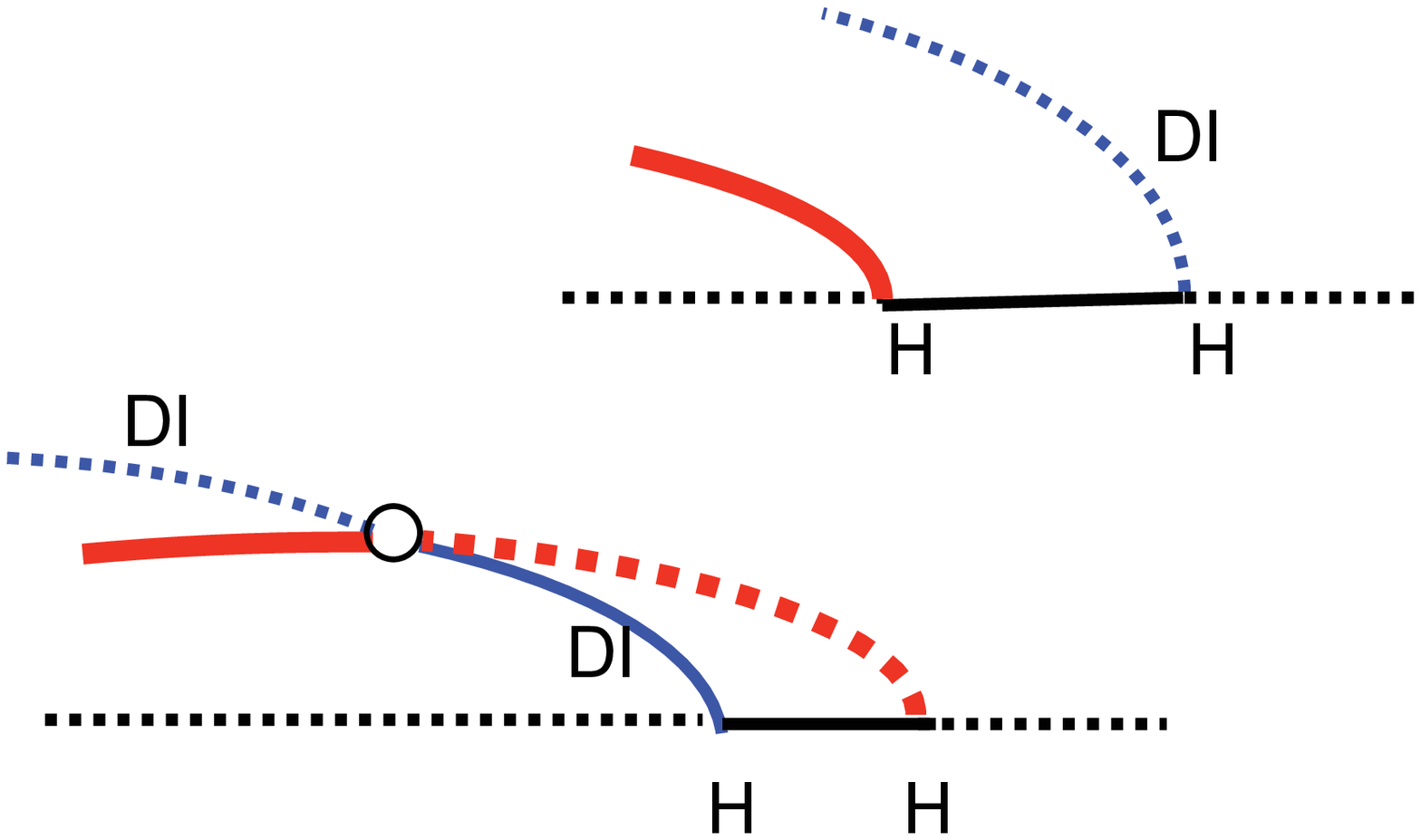}} 
\hspace{-1cm}
\subfloat[][Case (1c)] {\includegraphics[width=5cm]{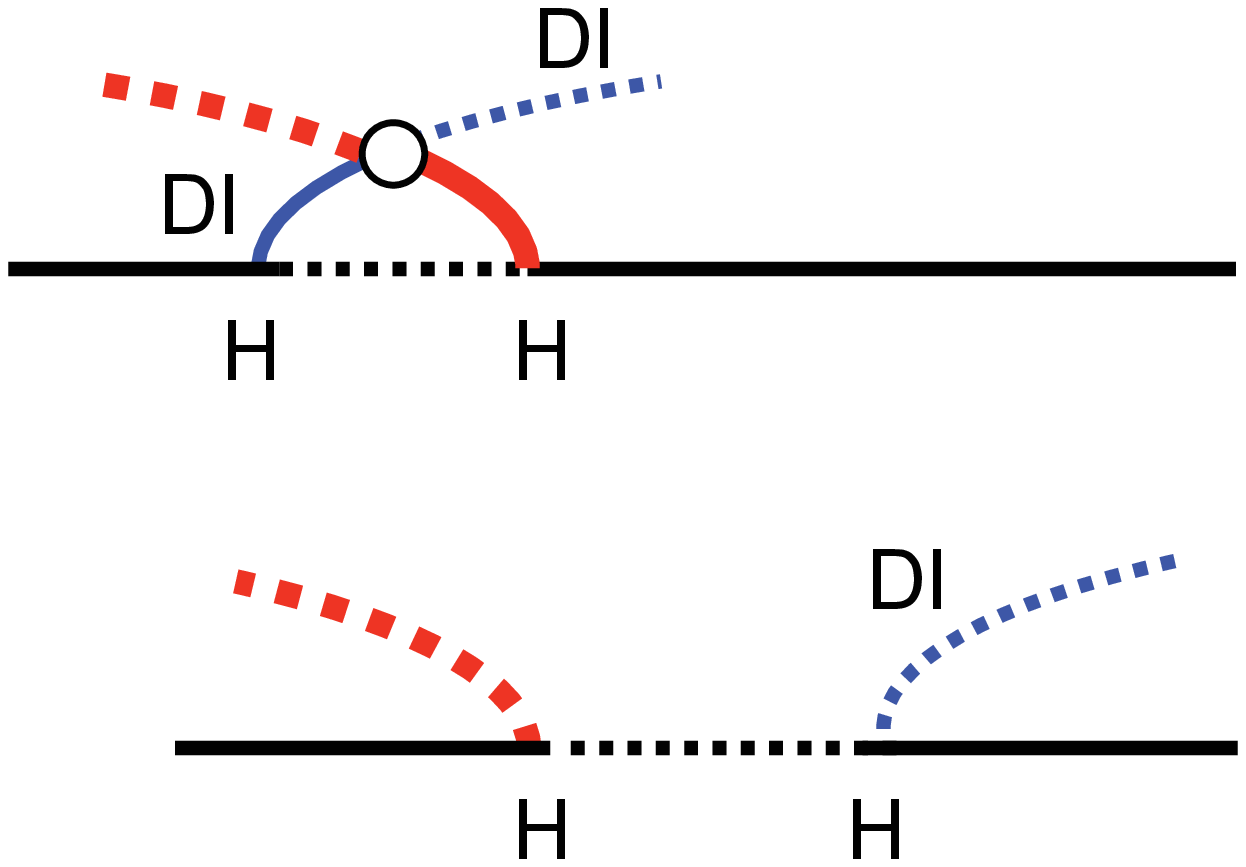}} \hspace{-1.5cm} \\
\hspace{-1.5cm}
\subfloat[][Case (2a)] {\includegraphics[width=5cm]{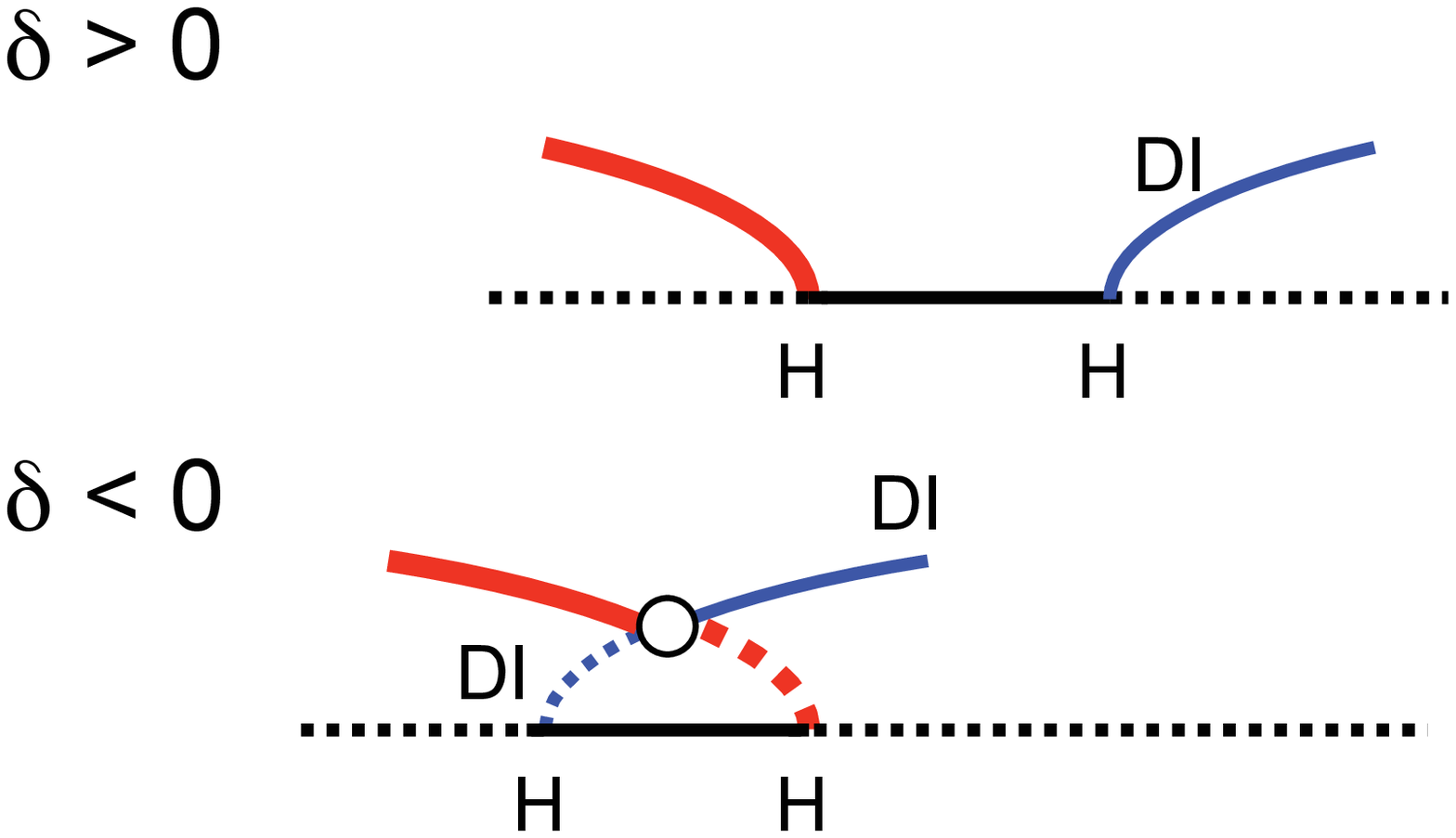}}  
\hspace{-1cm}
\subfloat[][Case (2b)] {\includegraphics[width=5cm]{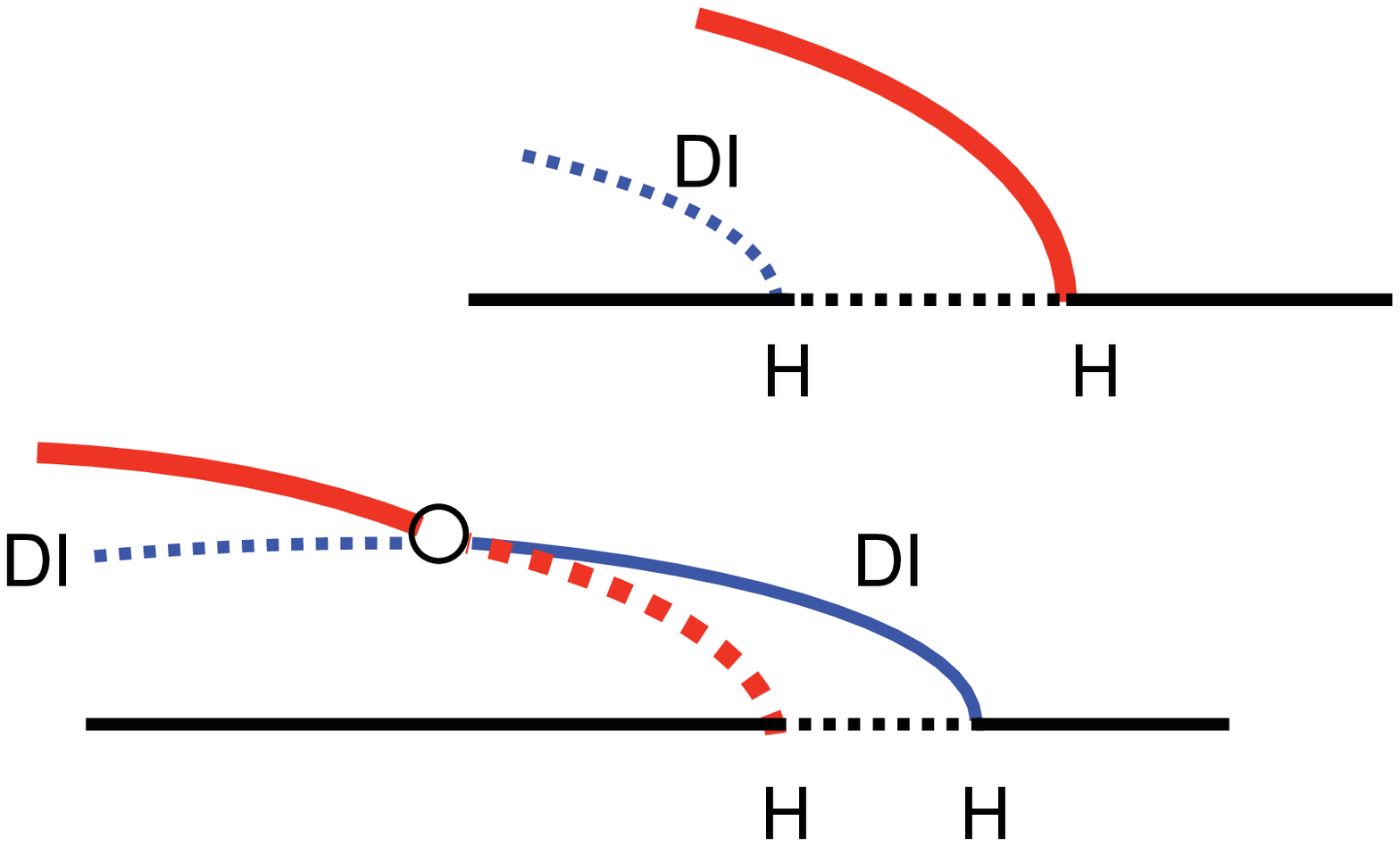}} 
\hspace{-1cm}
\subfloat[][Case (2c)] {\includegraphics[width=5cm]{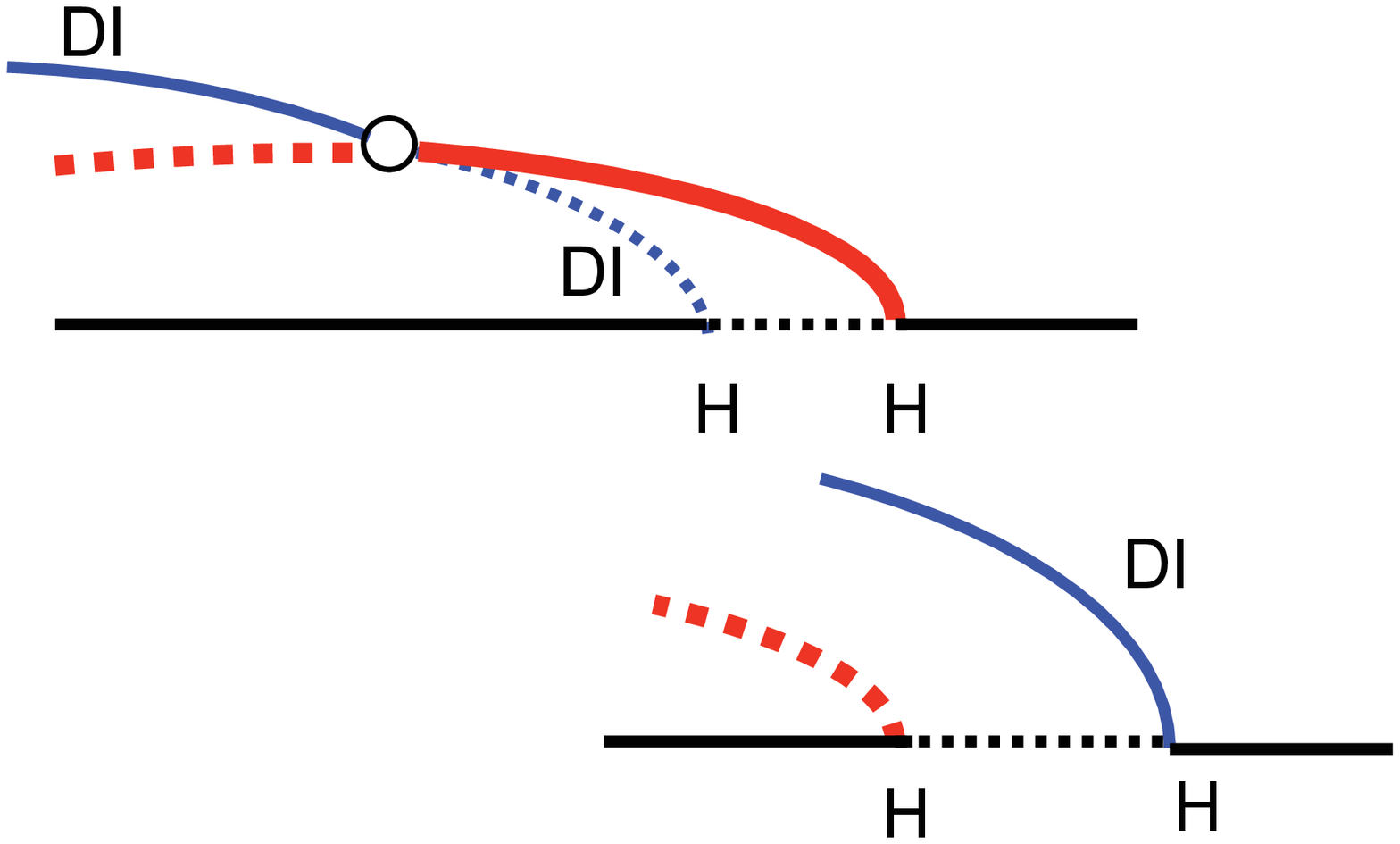}} \hspace{-1.5cm} \\
\caption{Bifurcation diagrams showing the amplitude of the periodic orbits vs. the parameter $\lambda$ for~\eqref{eq:factored} (see~\eqref{tab:cases} for description of cases).  Solid (dotted) lines indicate stable (unstable) solutions.  Hopf bifurcations are indicated by an H, and the transcritical bifurcation is indicated by an open dot.  The delay-induced periodic orbits are labeled DI, while the Pyragas orbits are unlabeled red curves.}
\label{fig:lambdavaries}
\end{figure}

For each subcase, we show the regions of existence of the Pyragas orbits and delay-induced orbits, their stability, and the stability of the trivial equilibrium in Fig.~\ref{fig:fourcases}.  Bifurcation diagrams showing the amplitude of the Pyragas and delay-induced periodic orbits as a function of the parameter~$\delta$ are shown in Fig.~\ref{fig:deltavaries}.  In both Cases (1) and (2) the Pyragas orbits are stabilized as soon as $\delta$ is increased beyond the transcritical bifurcation.  However, for Case (1), there is a smooth transition from the stable delay-induced orbit to the stable Pyragas orbit, whereas for Case (2), the zero solution can coexist stably with the delay-induced or Pyragas orbit, and hence hysteresis is expected.  In Fig.~\ref{fig:lambdavaries} we present bifurcation diagrams showing the amplitude of the periodic orbits vs.~the distinguished bifurcation parameter $\lambda$.  Six distinct cases are manifest, but, if $\delta > 0$, there is always a region in which the Pyragas orbits are stable in a neighborhood of $\lambda = 0$.  It is on the basis of this observation that we assert that Pyragas control can stabilize the subcritical branch of periodic orbits, provided $\gamma \neq 0$ and $\beta$ satisfies~\eqref{eq:restriction}.

\subsection{Connection with Singularity Theory}
The degenerate Hopf bifurcation~\eqref{eq:cmnormalformcomplex} can be related to a degenerate steady-state bifurcation problem with a $Z_{2}$ symmetry by focusing on the $\dot{r}$ equation.  This takes the form 
\begin{equation}
\dot{r} = s(u; \lambda)r, \hspace{0.5cm} u \equiv r^{2}, \nonumber
\end{equation}
with defining condition $s(0,0) = 0$ ensuring that a bifurcation of the $r = 0$ solution occurs at $\lambda = 0$.  The degenerate bifurcation of interest is defined by 
\begin{equation}
\frac{\partial s}{\partial \lambda}(0,0) = \frac{\partial s}{\partial u}(0,0) = 0. \nonumber
\end{equation}
This is codimension-two as a bifurcation problem, or a codimension-three phenomena (i.e. including $s(0,0) = 0$).  This bifurcation problem is analyzed using methods of singularity theory in the book by Golubitsky and Schaeffer (Chapter VI)~\cite{Golubitsky}.  They prove that it has the normal form 
\begin{equation}
\dot{r} = r \left(	a_{1}r^{4} + 2m\lambda r^{2} + a_{2}\lambda^{2} \right) \nonumber
\end{equation}
provided $m^{2} \neq a_{1} a_{2}$, where, by suitable rescaling, it is possible to set $a_{1} = \pm 1$ and $a_{2} = \pm 1$.  They analyze its universal unfolding
\begin{equation}
\dot{r} = r \left( a_{1}r^{4} + 2m\lambda r^{2} + a_{2}\lambda^{2} + a_{4} + 2a_{5}r^{2} \right). \nonumber
\end{equation}
Because we assume that the time delay $\tau$ coincides {\em exactly} with the period of the UPO, the unfolding parameter $a_{4} = 0$ in our problem.  Specifically, we have $a_{5} \propto \delta$, and we retain a term proportional to $\lambda \delta$ in place of $a_{4}$.  The singularity theory unfolding results are expected to apply directly if we were to consider deviations of $\tau$ from the period of the UPO.

\section{Center Manifold Reduction of the Delay Differential Equation}
\label{sec:cmreduction}
In this section, we use a center manifold reduction for delay differential equations to confirm that the cubic coefficient of the normal form~\eqref{eq:cmnormalformcomplex} is purely imaginary at $(\lambda,b_{0}) = (0,b_{c}^{0})$.  The theory is well-developed and is described thoroughly in, for example, \cite{OldHale,HaleLunel}.  In general, because the center manifold cannot be determined exactly, an approximation must be constructed, and this calculation can be facilitated by using a computer algebra program such as Maple\texttrademark \cite{SAC}.

We perform the reduction at $\lambda = 0$ so that $\tau = \tau(0) = 2\pi/\omega_{0}$.  Here we will focus on the simple case in which~\eqref{eq:origsetup} has no quadratic nonlinearities when Taylor-expanded about the origin.  We relegate the general case, in which quadratic nonlinearities are also present, to~Appendix~B.  We rewrite~\eqref{eq:origsetup} as 
\begin{equation}
\label{eq:delayeq}
\dot{\mathbf{x}}(t) = (\mathbf{J}_{0} - \mathbf{G}_{0})\mathbf{x}(t) + \mathbf{G}_{0}\mathbf{x}(t - 1) + \mathbf{f}(\mathbf{x}(t)), 
\end{equation}
where $\mathbf{x}(t) = \left(x,y,\mathbf{w} \right)^{T}$ and the matrices $\mathbf{J}_{0}$ and $\mathbf{G}_{0}$ are given by
\begin{equation}
\mathbf{J}_{0}= \frac{2\pi}{\omega_{0}} \left(	\begin{array}{cc} 
\begin{array}{cc}
0 & -\omega_{0} \\
\omega_{0} & 0 
\end{array}
 & \mathbf{O}_{(2,n-2)} \\ 
\mathbf{O}_{(n-2,2)} & \mathbf{D}(0)
\end{array}
\right), \nonumber
\end{equation}

\begin{equation}
\mathbf{G}_{0} =  \frac{2\pi}{\omega_{0}} \left(	
\begin{array}{cc} 
\begin{array}{cc}
b_{0}\cos \beta & -b_{0}\sin \beta \\
b_{0}\sin \beta  & b_{0}\cos \beta 
\end{array}
 & \mathbf{O}_{(2,n-2)} \\ 
\mathbf{O}_{(n-2,2)} & \mathbf{O}_{(n-2,n-2)} 
\end{array}
\right). \nonumber
\end{equation}
As in~\eqref{eq:origsetup}, $\mathbf{O}_{(i,j)}$ is an $i \times j$ zero matrix and $\mathbf{D}(0)$ is an $(n-2) \times (n-2)$ matrix in real Jordan normal form containing the decaying eigenvalues.  For the case we consider here, after Taylor-expanding about the origin, the vector field of nonlinear terms takes the form
\begin{equation}
\mathbf{f}(\mathbf{x}(t)) = \mathbf{f}_{3} + \dots,
\end{equation}
where 
\begin{equation}
\label{eq:F3}
\mathbf{f}_{3}  = \frac{2\pi}{\omega_{0}} \left( \begin{array}{ccc} 
 f_{14}x^3  + f_{15}x^2y + f_{16}xy^2 + f_{17}y^3 \\ 
f_{24}x^3 + f_{25}x^2y + f_{26}xy^2 + f_{27}y^3 \\
\mathbf{f}_{d4}x^3 + \mathbf{f}_{d5}x^2y + \mathbf{f}_{d6}xy^2 + \mathbf{f}_{d7}y^3 
\end{array} \right) + \dots 
\end{equation}
and (cubic) terms containing any of the $w_{i}$, $i = 1 \dots n-2$ have not been explicitly written because they do not contribute to the subsequent calculation.  The quantities $f_{14} \dots f_{27}$ are scalars and $\mathbf{f}_{d4} \dots \mathbf{f}_{d7}$ are vectors of dimension $n-2$. 

We follow \cite{SAC} in performing the center manifold reduction of~\eqref{eq:delayeq} to Hopf normal form at $\lambda = 0$.  In order to construct an appropriate phase space for the solutions of the delay differential equation, we define:  
\begin{equation}
\label{eq:newphasespace}
\mathbf{x}_{t}(\theta) \equiv \mathbf{x}(t + \theta),  \hspace{2mm} -1 \leq \theta \leq 0. \nonumber
\end{equation}
(Recall we rescaled time by the delay $\tau = 2\pi/\omega_{0}$, so that the delay time is fixed to be equal to $1$.)  We write \eqref{eq:delayeq} as a functional differential equation
\begin{equation}
\label{eq:FDE}
\dot{\mathbf{x}}(t) = L(\mathbf{x}_{t}) + \mathbf{f}(\mathbf{x}_{t})  \nonumber
\end{equation}
evolving in the Banach space $\mathcal{B} = C\left(\left[-1,0 \right], \mathbb{R}^{n} \right)$ \cite{HaleLunel}, 
where $L$ is a linear mapping defined by 
\begin{equation}
\label{eq:linearmapping}
L(\phi) = (\mathbf{J}_{0} - \mathbf{G}_{0}) \phi(0) + \mathbf{G}_{0} \phi(-1), 
\end{equation}
and $\mathbf{f}$ is a nonlinear functional defined by 
\begin{equation}
\label{eq:nonlinearfunctional}
\mathbf{f}(\phi) = \mathbf{f}(\phi(0)), 
\end{equation}
where $\phi \in \mathcal{B}$.  Here we have only $\phi(0)$ and $\phi(-1)$ on the right-hand sides of \eqref{eq:linearmapping} and \eqref{eq:nonlinearfunctional} because there is a single fixed delay that appears only in the linear terms of~\eqref{eq:delayeq}.  Linearizing \eqref{eq:FDE} about the trivial solution $\mathbf{x}(t) = 0$ we obtain
\begin{equation}
\label{eq:FDElin}
\dot{\mathbf{x}}(t) = L(\mathbf{x}_{t}). 
\end{equation}
As discussed in \cite{OldHale}, at the Hopf bifurcation point, the characteristic equation associated with~\eqref{eq:FDElin} has two purely imaginary roots.  We are interested in the case where the remaining (infinite number) of roots have negative real parts.  (The conditions to ensure this led to the restriction placed on $\beta$ determined in Section~\ref{sec:delayprob}.)  Hence, the solution space of~\eqref{eq:FDElin} can be decomposed as $\mathcal{B} = N  \oplus S$, where $N$ is a two-dimensional ``neutral" (center) eigenspace spanned by the solutions to~\eqref{eq:FDElin} corresponding to the eigenvalues with zero real part, and $S$ is the (infinite-dimensional) stable eigenspace.  

The goal of the decomposition is to determine the bases needed for deriving the two-dimensional equation that governs the dynamics on the center manifold.  When system~\eqref{eq:delayeq} contains only cubic nonlinearities, the equation can be obtained by calculating just two quantities associated with the linear problem, since the tangent plane approximation to the center manifold applies at leading order.  First we need a basis $\mathbf{\Phi}(\theta)$ for the center eigenspace $N$ of the linear problem, with $\theta \in [-1,0]$.  Second, we need a basis $\mathbf{\Psi}(\xi)$ for the center eigenspace of a linear problem dual to~\eqref{eq:FDElin}, with $\xi \in [0,1]$.  The system dual to~\eqref{eq:FDElin} is defined via the bilinear form given in \cite{OldHale}; the bilinear form is also used to normalize $\mathbf{\Psi}(\xi)$.  To summarize, the equation on the center manifold takes the form of a two-dimensional ordinary differential equation, which to cubic order is
\begin{equation}
\label{eq:udotcubic}
 \dot{\mathbf{u}} = \mathbf{H}\mathbf{u} + \mathbf{\Psi}(0)\mathbf{f}_{3} \left( \mathbf{\Phi}(0) \mathbf{u} \right)
\end{equation}
where 
\begin{equation}
\label{eq:Hmatrix}
\mathbf{H} = 
\left(\begin{array}{cc}
0 & 2\pi \\
-2\pi & 0 \\
\end{array} \right), \nonumber
\end{equation}

\begin{equation}
\label{eq:Phi0}
\mathbf{\Phi}(0)= \left(	\begin{array}{c}
\begin{array}{cc}
1 & 0 \\
0 & -1 \\
\end{array} \\
\mathbf{O}_{(n-2,2)} 
\end{array}
\right),
\end{equation}
and

\begin{equation}
\label{eq:Psi0}
\mathbf{\Psi}(0) = \left( \begin{array}{cc} 
\begin{array}{cc}
\left(A + B\cos \beta \right) & B\sin \beta \\
B\sin \beta & -\left(A + B\cos \beta \right)  
\end{array} & \mathbf{O}_{(2,n-2)} \\
\end{array} \right), 
\end{equation}
with 
\begin{align}
\label{eq:AB}
A &= \frac{\omega_{0}^{2}}{\omega_{0}^2 + 4\omega_{0}\pi b_{0}\cos \beta + 4\pi^2 b_{0}^{2}} \nonumber \\
B &= \frac{2 \pi \omega_{0} b_{0}}{\omega_{0}^2 + 4\omega_{0}\pi b_{0}\cos \beta + 4\pi^2 b_{0}^{2}}.
\end{align}

In complex form, \eqref{eq:udotcubic} becomes 
\begin{equation}
\label{eq:zdotcubic}
\dot{z} = 2 \pi i z + \frac{2\pi}{\omega_{0}}\left(A + Be^{-i \beta}  \right)F_{3}(z,\bar{z}).  
\end{equation}
Note that the only place the delay feedback parameters appear in~\eqref{eq:zdotcubic} is in the prefactor $\left(A + Be^{-i\beta}\right)$.  The real part of the coefficient of $|z|^2 z$ in $F_{3}$, which is the only term that survives the normal form transformation~\cite{GH}, takes the form
\begin{align}
\label{eq:krcubic}
k_{R} =& \left(A + B\cos \beta \right) c_{0} + \left(B \sin \beta \right)d_{0} \\
=& c_{0}\Big[ A + B (\cos \beta + \gamma  \sin \beta) \Big],\nonumber
\end{align}
where $\gamma = d_{0}/c_{0}$, and
\begin{align}
\label{eq:c0d0}
c_{0} &= \frac{3}{8}f_{14} + \frac{1}{8}f_{16} + \frac{1}{8}f_{25} + \frac{3}{8}f_{27} \\ 
d_{0} &= -\frac{3}{8}f_{17} -\frac{1}{8}f_{15} + \frac{1}{8}f_{26} + \frac{3}{8}f_{24}. \nonumber
\end{align}
The quantities $c_{0}$ and $d_{0}$ in~\eqref{eq:c0d0} are the real and imaginary parts, respectively, of the cubic coefficient in the uncontrolled problem restricted to the center manifold~\eqref{eq:nofbnormform} evaluated at $\lambda = 0$.  We can see that $k_{R} = 0$ when $A + B (\cos\beta + \gamma \sin \beta) = 0$.  Using~\eqref{eq:AB} we find that the value of $b_{0}$ for which this occurs is when 
\begin{equation}
b_{0} = b_{0}^{c} = \frac{-\omega_{0}}{2\pi \left( \cos \beta + \gamma \sin \beta \right)}. \nonumber
\end{equation}
Using center manifold theory, we have thus shown that the cubic coefficient of the normal form vanishes at $b_{0} = b_{0}^{c}$, which confirms the results based on the heuristic argument given in Section 3.

\section{Discussion}
We have shown that the branch of small amplitude UPOs created in a generic subcritical Hopf bifurcation from a stable equilibrium can be stabilized using an appropriate Pyragas-type feedback, provided that $\gamma \neq 0$ and $\beta$ satisfies~\eqref{eq:restriction}, where $\gamma$ is proportional to the imaginary part of the cubic coefficient of the Hopf normal form for the uncontrolled problem.  Specifically, we followed~\cite{PS} and considered  feedback only in the directions that are tangent to the center manifold at the bifurcation point.  As in~\cite{Fiedler}, the feedback gain matrix is parameterized by a gain amplitude, $b_0$, and a phase angle, $\beta$.  This choice of gain matrix reduces the problem of choosing $n^2$ components of a matrix, to that of choosing just the two parameters $b_0$ and $\beta$.  Moreover, we note that knowledge of $\gamma$ is already assumed for Pyragas control since it is needed to determine the time-delay for the feedback at leading order in $\lambda$.  No other information about the nonlinearities of the problem is required, making this a particularly simple and elegant control.

The behavior of the controlled system for $b_0$ near $b_0^c$  is governed by a highly degenerate Hopf bifurcation problem. In particular, {\it both} of the nondegeneracy conditions for a generic Hopf normal form are violated {\it simultaneously} : (a) the eigenvalues of the linearized problem do not cross the imaginary axis as the bifurcation parameter is varied, and (b) the real part of the cubic coefficient of the Hopf normal form, which determines whether the bifurcation is subcritical or supercritical, vanishes. We performed a center manifold reduction of the governing delay differential equations to prove this result in Section~\ref{sec:cmreduction}. The reason  this degeneracy may be surprising is that only \emph{two} parameters ($\lambda$ and $b_0$) are varied, and yet a codimension-three problem is obtained.  In Section~\ref{sec:delayprob} we demonstrated that this is due to the fact that the time-delay of the Pyragas feedback is fixed to coincide with the period of the original UPO.  A consequence of this restriction is that the existence properties for the bifurcating branch of targeted periodic orbits is unaffected by the feedback, although its stability is altered.  The transcritical bifurcation from classic bifurcation theory provides an apt analogy. This bifurcation typically occurs when there is an (equilibrium) solution whose existence is unaffected by the  value of the control parameter, {\it e.g.} some trivial equilibrium state ${\bf x}={\bf 0}$ that exists for all parameter values. For problems with such structure, the transcritical bifurcation supplants the saddle-node bifurcation as the generic steady state bifurcation problem~\cite{Golubitsky}. 

Just as one can unfold a transcritical bifurcation with imperfections, our degenerate Hopf bifurcation problem at $(\lambda,b_{0})=(0,b_0^c)$ could be unfolded further by considering detunings of the delay time $\tau$ from the period of the UPO. (This unfolding would also allow us to apply directly more of the singularity theory results for the degenerate bifurcation problem, developed by Golubitsky and Schaeffer~\cite{Golubitsky} and described in Section~\ref{sec:delayprob}.) We note that Just {\it et al.}~\cite{Just} have investigated bifurcations with varying $\tau$ in the setting of the Pyragas controlled subcritical Hopf normal form~\eqref{eq:normalform}.
They analyze a series of bifurcations in the three-parameter space $\tau$-$\lambda$-$b_0$ (in our notation). Since they allow $\tau$ to vary, the point $\lambda=0$, $\tau=2\pi$, $b_0=b_0^c$ in their case is a  codimension-three point, and there is a complicated sequence of bifurcations about this point.

The fundamental mechanism at work for the Pyragas control of UPOs arising from a subcritical Hopf bifurcation was first explained in the paper of Fiedler, {\it et al.}~\cite{Fiedler}. They showed that the feedback introduces  delay-induced instabilities of the bifurcating equilibrium, thereby altering its stability properties in a neighborhood of the original Hopf bifurcation. In this way, the equilibrium can change from being stable for $\lambda<0$ to being stable for $\lambda>0$, and then the original subcritical Hopf bifurcation (with increasing $\lambda$) can be converted to a supercritical Hopf bifurcation (with decreasing $\lambda$). This understanding suggests a number of interesting directions for generalizing Pyragas feedback for other bifurcations, where a similar simple control mechanism may work.

The Hopf bifurcation is just one of several bifurcation mechanisms by which UPOs can generically arise in dynamical systems.
As mentioned in the introduction, two  other generic mechanisms for creating UPOs in dynamical systems include homoclinic (or heteroclinic) global bifurcations and saddle-node bifurcations of limit cycles. The application of Pyragas-type feedback to stabilize UPOs arising in some of these situations has already been studied in the context of specific examples~\cite{Claire,Fiedlerfold}, and in both the saddle-node and heteroclinic case, the UPO is stabilized in a steady-state bifurcation with a delay-induced periodic orbit. In the heteroclinic case, the original problem was again of a higher dimension than the UPO. The gain matrix was chosen as in this paper, to be tangent to the two-dimensional manifold containing the UPO. We emphasise that this approach, studying the generic bifurcations which generate UPOs, leads to results which allow one to choose a gain matrix and predict the parameter region for which control can be achieved, with very little knowledge of the structure of the UPO.  Preliminary calculations attempting to stabilize a UPO arising from a generic homoclinic bifurcation have indicated that this is in fact \emph{not} possible, in distinct contrast to the other cases~\cite{Postlethwaite-hom}. 

An additional promising direction for generalizing Pyragas feedback for control of UPOs is to Hopf bifurcation problems with symmetry. A feature of such bifurcations, which arise naturally in pattern forming systems~\cite{CrawfordKnobloch}, is that they typically involve multiple complex conjugate pairs of eigenvalues crossing the imaginary axis simultaneously. Thus the results of this paper do not apply directly since we assumed that a {\it simple} complex conjugate pair crosses the imaginary axis at $\lambda=0$. Moreover, in equivariant Hopf bifurcation problems, a number of periodic orbits, distinguishable by their group of symmetry properties, may bifurcate simultaneously~\cite{Golubitsky2}. This group theoretic classification result suggests investigating feedback controls, analogous to Pyragas, that exploit the targeted symmetry properties (which may be spatial, temporal, or a combination of spatio-temporal symmetries) in such a way that the feedback vanishes on the targeted state only. The goal would be to use the feedback control to change stability properties of the bifurcating equilbrium, while naturally preserving the bifurcation to the targeted state.  Some first steps in this direction of controlling patterns arising in a Hopf bifurcation, using extensions of Pyragas control tailored to spatial symmetries (as well as temporal ones), were shown to work for certain plane wave solutions of the complex Ginzburg-Landau equation in regions of the Benjamin-Feir unstable regime~\cite{MontSilber,PS2}.  Further work on the stabilization of unstable patterns using these methods is ongoing.

\section{Acknowledgements}
 The authors thank Luis Mier-y-Ter\'{a}n-Romero for extensive discussions of center manifold reduction for delay differential equations and for assistance with \textsc{dde-biftool}~\cite{ddebiftool}.  They also thank Sue Ann Campbell for useful discussion of bifurcation analysis of delay differential equations and for sharing some of her Maple\texttrademark  code.  G.B. acknowledges support from NSF-RTG Grant (DMS-0636574) and the ARCS Foundation of Chicago.  She is grateful for the hospitality of the University of Auckland Mathematics Department during her visit as an NSF-EAPSI fellow.  C.M.P. acknowledges support from the University of Auckland Research Committee, and M.S. from the National Science Foundation (DMS-0709232).

 \begin{appendices}
\newcommand{\appsection}[1]{\let\oldthesection\thesection
  \renewcommand{\thesection}{Appendix \oldthesection}
  \section{#1}\let\thesection\oldthesection}


 \numberwithin{equation}{section}
\appsection{}
\setcounter{figure}{0}
\label{sec:appendixa}
Our analysis of the normal form for the degenerate Hopf bifurcation problem~\eqref{eq:cmnormalformcomplex} associated with $\lambda = 0, b_{0} = b_{0}^{c}$, presented in Section 3, is valid provided that certain nondegeneracy conditions are met. For instance we require that $\sigma_{3} \neq 0$ in~\eqref{eq:sigma} so that the coefficient of the linear term has real part proportional to $\lambda^2$ at $b_0=b_0^c$. Moreover, we assume that at $(\lambda, b_0)=(0,b_0^c)$, the real part of the coefficient of the quintic term in the normal form is nonzero. We do not perform a center manifold reduction of the delay differential equation to fifth order to determine this quintic coefficient since our results show that the important qualitative features are the same whether the quintic coefficient is positive or negative.  In order to allay any possible concerns that there is a hidden structure to the problem associated with Pyragas control that would force the real part of the quintic coefficient to also be zero at $(\lambda, b_0)=(0,b_0^c)$, we present a simple numerical example here for which we show that the coefficient is nonzero. We expect therefore that this nondegeneracy condition will be satisfied generically.

\begin{figure}[t!]
\centering
\includegraphics[height=6cm]{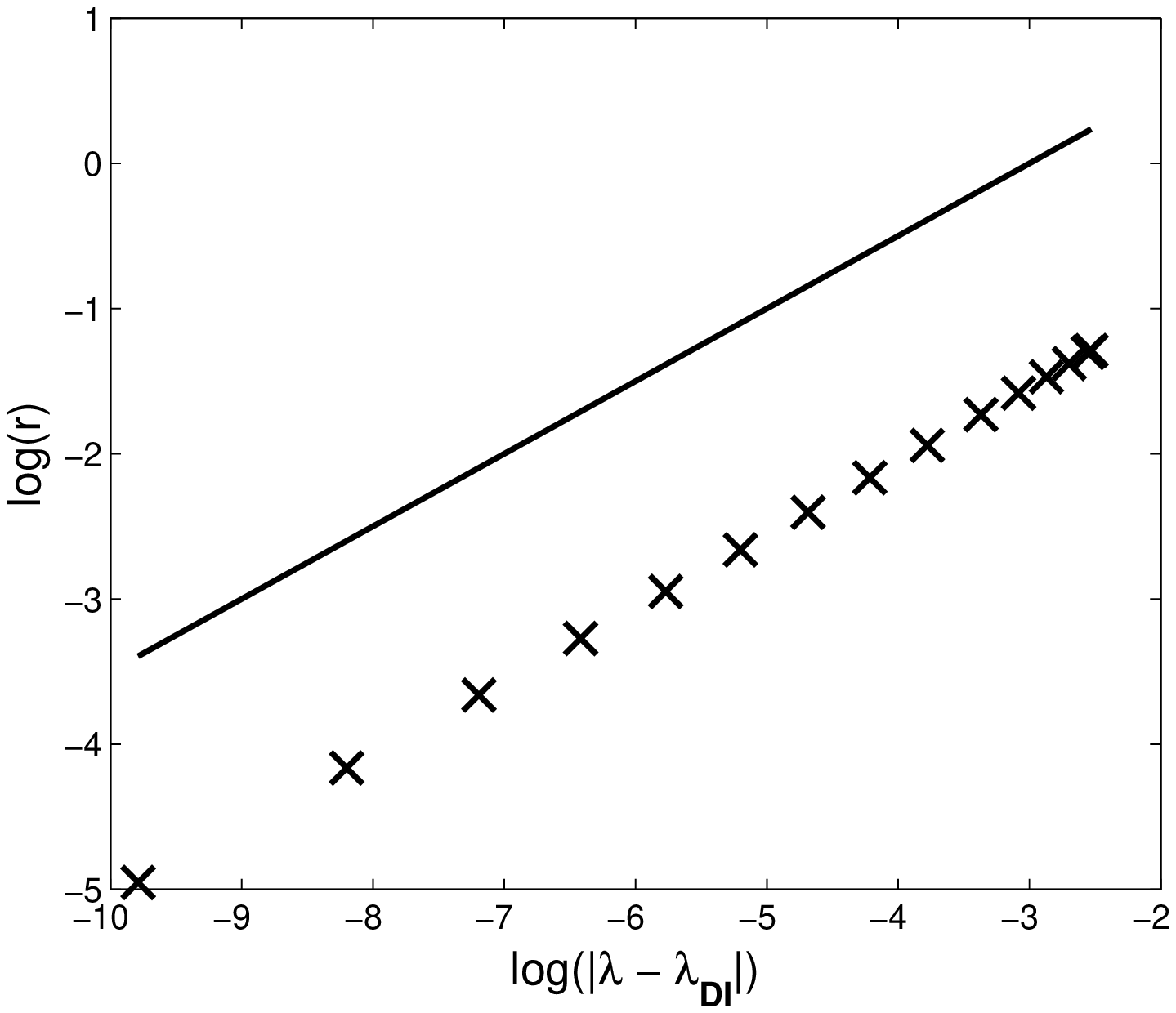}               
\caption{The figure shows $\log(r)$ vs.\ $\log|\lambda - \lambda_{DI}|$, indicated by x's, for the delay-induced periodic orbit computed from~\eqref{eq:hopfnfrescaled} using \textsc{dde-biftool}~\cite{ddebiftool} for $\delta \approx -0.0012$.  For comparison, a line with slope $0.5$ is plotted above the data points.}
\label{fig:scaling}
\end{figure}

Our numerical example is based on the Hopf normal form equation~\eqref{eq:normalform} with Pyaragas feedback control that was investigated by Fiedler {\it et al.}~\cite{Fiedler}.  After rescaling time by $\tau$ so that the delay is fixed, we have
\begin{equation}
\label{eq:hopfnfrescaled}
\dot{z}(t) = \tau (\lambda + i)z(t) + \tau(1 + i\gamma)|z(t)|^{2}z(t) + \tau b_{0}e^{i\beta}(z(t - 1) - z(t))
\end{equation}
with $\tau = 2\pi/(1 - \gamma \lambda)$.  
For our numerical example, we choose parameter values $\gamma = -10$, $\beta = \pi/8$.  Following the procedure in Section~\ref{sec:delayprob}, a linear stability analysis yields
\begin{align}
b_{0}^{c} &\approx 0.055 \nonumber \\
\sigma_{3} &\approx -247 \nonumber \\
\sigma_{4} &\approx -65, \nonumber
\end{align}
where the $\sigma_j$ appear in the expansion of the linear growth rate  $\sigma=\sigma_3\lambda^2+\sigma_4\lambda\delta+\cdots$ associated with perturbations of  the zero solution.

Since $\sigma_{3} \neq 0$, the normal form can, from~\eqref{eq:factored}, be written in the following factored form, valid in a neighborhood of $(\lambda,b_{0}) = (0,b_{0}^{c})$:
\begin{equation}
\label{eq:cmfactored}
\dot{r} = r (\lambda + r^{2})(\sigma_{3}\lambda + \sigma_{4}\delta + q_{R0}r^{2}). \nonumber
\end{equation}
We want to show that $q_{R0}\ne 0$, so we focus on the delay-induced periodic orbits with amplitude $r$ satisfying
\begin{equation}
\label{eq:delayind}
\sigma_{3}\lambda + \sigma_{4}\delta + q_{R0}r^{2} = 0.
\end{equation}
For a fixed $\delta$, the value of $\lambda$ at which a Hopf bifurcation to a delay-induced periodic orbit occurs, which we denote by $\lambda_{DI}(\delta)$, is defined by 
\begin{equation}
\sigma_{3}\lambda_{DI}(\delta) + \sigma_{4}\delta = 0. \nonumber
\end{equation}
Then (\ref{eq:delayind}) becomes
\begin{equation}
\sigma_{3}\left( \lambda - \lambda_{DI}(\delta) \right) + q_{R0}r^{2} = 0, \nonumber
\end{equation}
where 
\begin{equation}
\lambda_{DI}(\delta) = \frac{\sigma_{4}\delta}{-\sigma_{3}}\approx -0.264\ \delta. \nonumber
\end{equation}

We followed the branch of delay-induced periodic orbits in a neighborhood of $\lambda=\lambda_{DI}$ for a sequence of values of $\delta$ approaching  $\delta= 0$ using the bifurcation package
\textsc{dde-biftool}~\cite{ddebiftool}.  If $q_{R0}\ne 0$ and $\lambda$ sufficiently close to $\lambda_{DI}$, then we expect the amplitude $r$ of the delay-induced branch to scale as $\sqrt{|\lambda-\lambda_{DI}|}$.  In Figure~\ref{fig:scaling} we graph $\log(r)$ vs. $\log |\lambda-\lambda_{DI}|$ for $\delta \approx -0.0012$, and find the square root dependence of  $r$ on $|\lambda - \lambda_{DI}|$, which is consistent with $q_{R0} \neq 0$.  If $q_{R0} = 0$, then we would not expect this scaling.

\appsection{}
\label{sec:appendixb}
This Appendix presents the center manifold reduction for the case in which system~\eqref{eq:origsetup} contains quadratic nonlinearities, as well as cubic.  The goal is the same: to determine the two-dimensional equation that governs the dynamics on the center manifold.  However, because the tangent plane approximation to the center manifold no longer applies, we must now approximate the center manifold. 

We begin by rewriting \eqref{eq:origsetup} in the form \eqref{eq:delayeq}.  The vector field of nonlinear terms can be expanded as: 
\begin{equation}
\mathbf{f}(\mathbf{x}(t)) = \mathbf{f}_{2} + \mathbf{f}_{3} + \dots \nonumber
\end{equation}
where $\mathbf{f}_{3}$ is given by~\eqref{eq:F3} and 
\begin{equation}
\mathbf{f}_{2} = \frac{2\pi}{\omega_{0}}  \left( \begin{array}{ccc} 
 f_{11}x^2  + f_{12}xy + f_{13}y^2 + \sum_{i=1}^{n-2} \left( g_{1i}xw_{i} + k_{1i}yw_{i} \right)\\ 
f_{21}x^2 + f_{22}xy + f_{23}y^2 + \sum_{i=1}^{n-2} \left( g_{2i}xw_{i} + k_{2i}yw_{i} \right)\\
\mathbf{f}_{d1}x^2 + \mathbf{f}_{d2}xy + \mathbf{f}_{d3}y^2 + \sum_{i=1}^{n-2} \left( \mathbf{g}_{di}xw_{i} + \mathbf{k}_{di}yw_{i} \right)
\end{array} \right) + \dots \nonumber
\end{equation}
Here we have neglected to write terms quadratic in $w_{i}$, since they don't enter the subsequent calculation. 

With the bases $\mathbf{\Phi}(\theta)$ and $\mathbf{\Psi}(\theta)$ already given by \eqref{eq:Phi0} and \eqref{eq:Psi0}, respectively, the final quantity to be calculated is an approximation to the center manifold, which will be denoted by $\mathbf{h}(\mathbf{u}, \theta)$.  To obtain terms up to cubic order for the evolution equation on the center manifold, we need only consider terms up to quadratic order in the equation for $\mathbf{h}(\mathbf{u},\theta)$, that is, we write
\begin{equation}
\mathbf{h}_{2}(\theta, \mathbf{u}) = 
\left( \begin{array}{c}
h_{1\_11}(\theta)u^{2}_{1} + h_{1\_12}(\theta)u_{1}u_{2} + h_{1\_22}(\theta)u^{2}_{2} \\
h_{2\_11}(\theta)u^{2}_{1} + h_{2\_12}(\theta)u_{1}u_{2} + h_{2\_22}(\theta)u^{2}_{2} \\
\mathbf{h}_{d\_11}(\theta)u^{2}_{1} + \mathbf{h}_{d\_12}(\theta)u_{1}u_{2} + \mathbf{h}_{d\_22}(\theta)u^{2}_{2} \end{array} \right), \nonumber
\end{equation}
where  $h_{1\_11}(\theta), h_{1\_12}(\theta), \ldots h_{2\_22}(\theta)$ are scalar functions of $\theta \in [0,1]$ and $\mathbf{h}_{d\_11}(\theta)$, $\mathbf{h}_{d\_12}(\theta)$, and $\mathbf{h}_{d\_22}(\theta)$ are $(n-2)$-dimensional vector-valued functions.  As shown in \cite{SAC}, $\mathbf{h}_{2}$ must satisfy the equation 
\begin{equation}
\label{eq:diffeqns} 
\frac{\partial \mathbf{h}_{2}}{\partial \theta} = \frac{\partial \mathbf{h}_{2}}{\partial \mathbf{u}}(\theta,\mathbf{u})\mathbf{H}\mathbf{u} + \mathbf{\Phi}(\theta)\mathbf{\Psi}(0)\mathbf{f}_{2}(\mathbf{\Phi}(\theta)\mathbf{u}), 
\end{equation}
subject to the boundary condition
\begin{equation}
\label{eq:bcs} 
\frac{\partial \mathbf{h}_{2}}{\partial \mathbf{u}} \Big|_{\theta = 0} \mathbf{H} \mathbf{u} + \mathbf{\Phi}(0)\mathbf{\Psi}(0)\mathbf{f}_{2}(\mathbf{\Phi(\theta)}\mathbf{u}) = L(\mathbf{h}_{2}(\theta,\mathbf{u})) + \mathbf{f}_{2}(\mathbf{\Phi}(\theta)\mathbf{u}),
\end{equation}
where $\mathbf{H}$ is given by~\eqref{eq:Hmatrix}.  Solving \eqref{eq:diffeqns}, \eqref{eq:bcs} for the coefficients of the first two rows of $\mathbf{h}_{2}$ yields
\begin{equation}
\mathbf{h}_{2}(0) = \frac{1}{3\omega_{0}} \begin{pmatrix} (1 - A - B\cos\beta)\mathbf{I}_{3} & -(B\sin\beta) \mathbf{I}_{3} \\  (B\sin\beta)\mathbf{I}_{3} & (1 - A - B\cos\beta )\mathbf{I}_{3} \end{pmatrix} \mathbf{v}, 
\end{equation}
where $\mathbf{I}_{3}$ is the $3 \times 3$ identity matrix, $A,B$ are given by~\eqref{eq:AB} and 
\begin{equation}
\mathbf{v} = \begin{pmatrix}  - f_{12} - f_{21} - 2f_{23}  \\  - 2f_{11} + 2f_{13} - f_{22} \\ f_{12} - 2f_{21} - f_{23} \\ f_{11} + 2f_{13} - f_{22} \\ f_{12} - 2f_{21} + 2f_{23} \\ 2f_{11} + f_{13} + f_{22} \end{pmatrix}. 
\end{equation}
To determine the vectors $\mathbf{h}_{d \_11}(0)$, $\mathbf{h}_{d \_12}(0)$, and $\mathbf{h}_{d \_22}(0)$ we solve the following system of equations: 
\begin{equation}
\mathbf{M} \left( \begin{array}{c} \mathbf{h}_{d\_11}(0) \\  \mathbf{h}_{d\_12}(0) \\ \mathbf{h}_{d\_22}(0) \end{array} \right) = 
 \left( \begin{array}{c} \mathbf{f}_{d1}(0) \\  \mathbf{f}_{d2}(0) \\  \mathbf{f}_{d3}(0) \end{array} \right), \nonumber
\end{equation}
where 
\begin{equation}
 \mathbf{M} = \left( \begin{array}{ccc}
-\mathbf{D}(0) & -\omega_{0} \mathbf{I}_{n-2} &  \mathbf{O}_{n-2} \\
2 \omega_{0} \mathbf{I}_{n-2} & -\mathbf{D}(0) &  -2 \omega_{0} \mathbf{I}_{n-2} \\
 \mathbf{O}_{n-2} & \omega_{0} \mathbf{I}_{n-2} & -\mathbf{D}(0) \end{array} \right) \nonumber
\end{equation}
is a matrix of size $3(n-2) \times 3(n-2)$, $\mathbf{D}(0)$ is the real $(n-2) \times (n-2)$ matrix defined in~\eqref{eq:origsetup} and evaluated at $\lambda = 0$, $\mathbf{O}_{n-2}$ is a $(n-2) \times (n-2)$ zero matrix, and $\mathbf{I}_{n-2}$ is the $(n-2) \times (n-2)$ identity matrix. 

Finally, the equation on the center manifold is given by 
\begin{equation}
 \dot{\mathbf{u}} = \mathbf{H} \mathbf{u} + \mathbf{\Psi}(0)\big( \mathbf{f}_{2} \left( \mathbf{\Phi}(0)\mathbf{u}\right) + \nabla \mathbf{f}_{2}\left( \mathbf{\Phi}(0)\mathbf{u}\right)\mathbf{h}_{2}(0) + 
 \mathbf{f}_{3} \left( \mathbf{\Phi}(0)\mathbf{u}\right)  \big), \nonumber
\end{equation}
or in complex form,
\begin{equation}
\dot{z} = 2\pi i z + \frac{2 \pi}{\omega_{0}}\left(A + Be^{-i\beta} \right)F_{2,3}(z,\bar{z}), \nonumber 
\end{equation}
where $F_{2,3}(z,\bar{z})$ contains both quadratic and cubic terms proportional to $z^{2}$, $|z|^{2}$, $\bar{z}^{2}$, $z^{3}$, etc.
After performing a near identity transformation \cite{GH}, the real part of the cubic coefficient $|z|^{2}z$ takes the form
\begin{eqnarray}
\label{eq:krquadratic}
k_{R} &=& \left(A + B\cos \beta \right) c_{0} + \left(B \sin \beta \right)d_{0} \\
	&=& c_{0}\Big[ A + B ( \cos \beta + \gamma  \sin\beta) \Big], \nonumber
\end{eqnarray}
with $\gamma = d_{0}/c_{0}$.  Once again, from equation \eqref{eq:krquadratic} we can see that $k_{R} = 0$ when 
\begin{equation}
b_{0} = b_{0}^{c} = \frac{-\omega_{0}}{2\pi (\cos\beta + \gamma \sin\beta)}. \nonumber
\end{equation}
For completeness, we find
\begin{equation}
\begin{array}{ccl}
c_{0} &=& \frac{3}{8}f_{14} + \frac{1}{8}f_{16} + \frac{1}{8}f_{25} + \frac{3}{8}f_{27} + \frac{1}{4 \omega_{0}}f_{13}f_{23} - \frac{1}{4 \omega}f_{11}f_{21} \\
 & &+\frac{1}{8 \omega_{0}}f_{11}f_{12} + \frac{1}{8 \omega_{0}}f_{12}f_{13} - \frac{1}{8 \omega_{0}}f_{21}f_{22} - \frac{1}{8 \omega_{0}}f_{23}f_{22} \\
 & &+\frac{3}{8}\sum_{i=1}^{n-2}g_{1i}h_{di\_11}(0) - \frac{1}{8}\sum_{i=1}^{n-2}\left( g_{2i}h_{di\_12}(0) -k_{2i}h_{di\_11}(0) \right) \\
 & &+\frac{1}{8}\sum_{i=1}^{n-2}\left(g_{1i}h_{di\_22}(0) - k_{1i}h_{di\_12}(0) \right) + \frac{3}{8}\sum_{i=1}^{n-2}k_{2i}h_{di\_22}(0) \\
 & &  \\
 d_{0} &=& \frac{3}{8}f_{24} + \frac{1}{8}f_{26} - \frac{1}{8}f_{15} - \frac{3}{8}f_{17} + \frac{1}{24 \omega_{0}}f_{12}f_{21} - \frac{1}{24 \omega_{0}}(f_{12})^2 \\
 & & -\frac{1}{24 \omega_{0}}(f_{22})^2 + \frac{1}{24 \omega_{0}}f_{13}f_{22} -  \frac{1}{24 \omega_{0}}(f_{22})^2 + \frac{1}{24 \omega_{0}}f_{13}f_{22} \\
& & - \frac{5}{12 \omega_{0}}f_{21}f_{23} - \frac{5}{12 \omega_{0}}f_{11}f_{13} - \frac{5}{12 \omega_{0}}(f_{21})^2 - \frac{5}{12 \omega_{0}}(f_{13})^2 \\
& & - \frac{1}{6 \omega_{0}}(f_{23})^2 - \frac{1}{6 \omega_{0}}(f_{11})^2  + \frac{5}{24 \omega_{0}}f_{11}f_{22} + \frac{5}{24 \omega_{0}}f_{23}f_{12} \\
& & +\frac{3}{8}\sum_{i=1}^{n-2}g_{2i}h_{di\_11}(0) + \frac{1}{8} \sum_{i=1}^{n-2}\left(g_{1i}h_{di\_12}(0) - k_{1i}h_{di\_11}(0)\right) \\
& & + \frac{1}{8}\sum_{i=1}^{n-2}\left(g_{2i}h_{di\_22}(0) - k_{2i}h_{di\_12}(0)\right) - \frac{3}{8}\sum_{i=1}^{n-2}k_{2i}h_{di\_22}(0), \nonumber
\end{array}
\end{equation}
which are the precisely the coefficients $c(0), d(0)$ in the uncontrolled Hopf normal form~\eqref{eq:nofbnormform}. 
 \end{appendices}



\end{document}